\documentclass[%
 reprint,
superscriptaddress,
 amsmath,amssymb,
 aps,
]{revtex4-1}
\usepackage[colorlinks,linkcolor=blue, citecolor=blue, urlcolor=blue]{hyperref}
\usepackage{graphicx}
\usepackage{dcolumn}
\usepackage{bm}
\usepackage{subfigure}
\usepackage{multirow}
\usepackage{float}
\usepackage{subfigmat}
\newcommand\vect[1]{\bm{#1}}

\newcommand\refeq[1]{equation \eqref{#1}}
\newcommand\reffig[1]{Fig. \ref{#1}}

\begin{document}

\preprint{APS/123-QED}

\title{Snowflake growth in three dimensions using phase field  modelling}%

\author{G. Demange}
 \affiliation{GPM, UMR CNRS 6643, University of Rouen, 76575 Saint \'Etienne du Rouvray, France}
\author{H. Zapolsky}
 \affiliation{GPM, UMR CNRS 6643, University of Rouen, 76575 Saint \'Etienne du Rouvray, France}
\author{R. Patte}%
 \affiliation{GPM, UMR CNRS 6643, University of Rouen, 76575 Saint \'Etienne du Rouvray, France}
\author{M. Brunel}
 \affiliation{CORIA UMR 6614, University of Rouen, Avenue de l'universit\'e BP 12, 76801 Saint \'Etienne du Rouvray, France}
\date{\today}

\begin{abstract}

Snowflake growth provides us with a fascinating example of spontaneous pattern formation in nature. Attempts to understand this phenomenon have led to important insights in non-equilibrium dynamics observed in various active scientific fields, ranging from pattern formation in physical and chemical systems, to self-assembly problems in biology. Yet, very few models currently succeed in reproducing the diversity of snowflake forms in three dimensions, and the link between model parameters and thermodynamic quantities is not established. Here, we report a modified phase field model that describes the subtlety of the ice vapour phase transition, through anisotropic water molecules attachment and condensation, surface diffusion, and strong anisotropic surface tension, that guarantee the anisotropy, faceting and dendritic growth of snowflakes. We demonstrate that this model reproduces the growth dynamics of the most challenging morphologies of snowflakes from the Nakaya diagram. We find that the growth dynamics of snow crystals matches the selection theory, consistently with previous experimental observations.
\end{abstract}

\pacs{Valid PACS appear here}
\maketitle

Snowflake growth in supersaturated atmosphere is one of the most familiar, and at the same time scientifically challenging physical phenomena \cite{PhysRevLett.108.193003}. Beyond aesthetic fascination for their symmetric shape, snow crystals provide us with a unique example of self-patterning systems \cite{ball2016material}. Early experiments on artificial snow crystals in cold chamber led by Nakaya \cite{nakaya1938preliminary}, revealed this fundamental phase transition resulted in a wide manifold of patterns, exclusively determined by supersaturation and temperature. This dependency was later formalized in the meteorological classification of Magono \cite{magono1966meteorological}, and the snow crystals morphology diagram of Nakaya \cite{nakaya1951formation}.

Despite this clear experimental picture, the underlying physical rules remain thoroughly debatable \cite{libbrecht2005physics}, as evidenced by the numerous models proposed to understand the variety of snowflake shapes, such as the surface diffusion model of Mason et al. \cite{mason1992snow}, the quasi-liquid layer approach of Lacmann et al. \cite{kuroda1982growth}, and the layer nucleation rates theory by Nelson \cite{nelson2001growth}. 

Alternatively, many simulation methods were developed to reproduce the growth dynamics of snowflakes. First step toward the comprehension of ice crystal growth was provided by molecular dynamics simulations \cite{nada1996anisotropic,furukawa1997anisotropic,benet2016premelting}. Unfortunately, such simulations are still confined to space and time scales by several orders inferior to snowflake characteristic scales \cite{magono1966meteorological}. Significant results were also achieved using the mesoscopic approach, such as the cellular automata model of Gravner and Griffeath \cite{gravner2009modeling}. Though this model remarkably describes the morphology of snowflakes, the numerous parameters of this model can hardly be related to physical quantities.


From this perspective, a 3D sharp interface model was developed by Barrett et al. \cite{barrett2012numerical}. Different snowflake morphologies were simulated. Nevertheless, the side branching \cite{libbrecht2005physics,libbrecht2006ken}, surface markings \cite{furukawa1993three,saito1996statistical,sazaki2010elementary}, and coalescence \cite{Kikuchi1998169} of ice crystals could not be reproduced in this  framework. Besides, only small supersaturations were prospected. Consequently, only the bottom of the Nakaya diagram was explored. This limitation comes from the Laplacian approximation framework, and the numerical cost of interface parametrization \cite{singer2008phase}.  The phase field model has the decisive advantage to overcome explicit tracking of the sharp interface, by spreading it out over a small layer \cite{singer2008phase}. As a matter of fact, it has become standard to simulate the dendritic growth in alloy solidification \cite{karma1996phase,ramirez2004phase,cartalade2016lattice}. However, despite a prototype attempt by Barret et al. in \cite{barrett2013stable}, the phase field approach was never used to simulate snow crystal growth in three dimensions. Indeed, until recently, it was assumed that the phase field approach was unable to reproduce facet formation and destabilization \cite{kobayashi1993modeling}. Yet, Debierre and Karma suggested in \cite{debierre2003phase}, that  phase field could mimic faceting using highly anisotropic surface tension.


In this paper, we report the simulations of snow crystals growth in three dimensions, using a modified phase field model. A new surface tension anisotropy function accounting for the 6-fold horizontal and 2-fold vertical symmetry of snowflakes was derived.  A supplementary anisotropy function, and anisotropic diffusion terms were also included to simulate the vertical anisotropy of snowflakes \cite{libbrecht2005physics}. To mimic faceting in snowflakes, the 2D regularization algorithm of Eggleston et al. \cite{eggleston2001phase} was extended to three dimensions. As a result, the model reproduces the growth of the main snowflake morphologies of the Nakaya diagram, varying only four phenomenological parameters. It is shown that these parameters can be related to  physical quantities. Simulated snowflakes show excellent agreement with experimental observations \cite{furukawa1993three,saito1996statistical,libbrecht2006ken,sazaki2010elementary,libbrecht2012toward}. Their growth satisfies the microscopic solvability theory \cite{langer1978evidence,langer1978theory,amar1993theory,brener1996three}, consistently with experiments \cite{libbrecht2002electrically}. 


Snowflake growth in supersaturated water vapour was simulated using a phase field approach in three dimensions, based on a methodology developed in \cite{karma1996phase}. In this model, two coupled variables $\phi$ and $u$ are considered. $\phi$ is an order parameter referring to the ice ($+1$) and vapour ($-1$) phases. The ice/vapour interface is described by a continuous variation of $\phi$, connecting $-1$ and $+1$. $u=(c-c_{\text{sat}}^I)/c_{\text{sat}}^I$ is the reduced supersaturation of water vapour, where $c_{\text{sat}}^I(T)$ is the saturation number density of vapour above ice, at temperature $T$. Then, the growth kinetics of snowflakes is governed by two non conservative phase field equations. Their adimensionalized form  is given by:

\begin{align}
A(\vect{n})^2\partial_t \phi&= -f'(\phi) +  \lambda B(\vect{n})g'(\phi)u \label{1}\\
&+\frac{1}{2}\nabla_{\Gamma} \cdot\left(|\nabla\phi|^2 \frac{\partial \left[A(\vect{n})^2\right]}{\partial \nabla\phi} + A(\vect{n})^2\nabla_{\Gamma} \phi\right)\nonumber\\
\partial_t u &=\widetilde{D}\nabla_{\Gamma}\cdot(q(\phi)\nabla_{\Gamma}) u-\frac{L_{\text{sat}}}{2} B(\vect{n})\partial_t \phi,\label{2}
\end{align}
where space and time are scaled by the interface width $W_0$, and the relaxation time $\tau_0$ respectively. In \refeq{1}, the double well potential $f(\phi)=-\phi^2/2+\phi^4/4$ is the free energy density of the ice/vapour system, at temperature $T$, and saturation concentration $c=c_{\text{sat}}^I$. The second term in \refeq{1} accounts for the coupling between $u$ and $\phi$, and promotes ice phase growth in supersaturated atmosphere, where $c>c_{\text{sat}}^I$. Formally, it corresponds to the first order term in the Taylor expansion of the bulk potential, for $c$ in the neighbourhood of $c_{\text{sat}}^I$. The coupling constant $\lambda$ can thus be computed by $\lambda=\left.\partial_c f\right|_{c_{\text{sat}}}c_{\text{sat}}/(30 H)$, where $H$ is the free energy barrier between vapor and solid phases. $g'(\phi)=(1-\phi^2)^2$ is an interpolation function introduced in \cite{karma1996phase}.  Its form allows to keep the bulk potential minima at $\phi=\pm 1$ for any $u$. To describe the strong anisotropy of snowflake growth along the vertical axis \cite{libbrecht2005physics}, we propose to introduce the anisotropy function $B(\vect{n})=\sqrt{n_x^2+n_y^2+\Gamma^2 n_z^2}$. Here, $\vect{n}=-\nabla \phi/|\nabla \phi|$ is the unit normal vector of $\phi$. The parameter $\Gamma>0$ governs the preference between horizontal and vertical growth, called the primary habit of snowflakes. It can be empirically related to the temperature \cite{nelson2001growth}. The addition of the first two terms in \refeq{1} corresponds to the anisotropic thermodynamic driving force.

The last two terms in \refeq{1} describe the ice/vapour interface formation and propagation \cite{karma1996phase}, where $A(\vect{n})$ is the surface tension anisotropy function. In this work, a new expression of the anisotropy function was derived: $A(\vect{n})=1+\epsilon_{xy}\cos(6\theta)+\epsilon_{z}\cos(2\psi)$, where $\theta=\arctan(n_y/n_x)$ and $\psi=\arctan(\sqrt{n_x^2+n_y^2}/n_z)$ are the polar and azimuthal angles respectively. It accounts for both the horizontal 6-fold symmetry, and the vertical planar symmetry  of snowflakes. $\epsilon_{xy}$ and $\epsilon_{z}$ are anisotropy constants.

The \refeq{2} describes the one-sided diffusion of water molecules in vapour, and vapour condensation on ice. Diffusion is controlled by two quantities: the function $q(\phi)=1-\phi$, which prohibits diffusion within ice, and the reduced diffusion coefficient $\widetilde{D}=D\tau_0/W_0^2$, where $D$ is the diffusion coefficient. The second term in \refeq{2} accounts for conversion of vapour into ice. $L_{\text{sat}}$ can thus be interpreted as the rate of water depletion in vapour, via molecule attachment at the interface. In this study, $L_{\text{sat}}$ was treated as a numerical parameter, and its setting was conditioned by the choice of $\lambda$. Here, $L_{\text{sat}}\sim 1$, to allow a sufficiently fast crystal growth. Obviously, attachment kinetics limited growth enabling snowflake faceting is lost. Faceting is yet recovered through the highly anisotropic interface, as suggested by Debierre and Karma in \cite{debierre2003phase}. To account for the anisotropy of the water molecule attachment kinetics on snow crystals \cite{nelson2001growth,kuroda1982growth}, the anisotropy function $B(\vect{n})$ is introduced in the second term of \refeq{2} as well. The vertical/horizontal growth preference is also considered through anisotropic diffusion \cite{furukawa1997anisotropic}, $\nabla_{\Gamma}=(\partial_x,\partial_y,\Gamma \partial_z)$. 

Constants $\lambda$, $W_0$, $\tau_0$ and $\widetilde{D}$ are entangled by the asymptotic analysis mapping the phase field model to Stefan sharp interface model, as shown in \cite{karma1996phase}. This sets $\widetilde{D}=0.6267 \lambda$,  and $W_0=d_0 \lambda/0.8839$, where $d_0$ is the isotropic capillarity length. This analysis also links the anisotropic interface width $W_0 A(\vect{n})$ to the characteristic time of interface propagation $\tau_0 A(\vect{n})^2$, where $\tau_0=0.6267 \lambda  W_0^2$.

The coupling constant was set to $\lambda=3.0$ as in \cite{ramirez2004phase}. The horizontal anisotropy  constant was set to $\epsilon_{xy}=0.1$ for vertical growth ($\Gamma\geq 1$), and $\epsilon_{xy}=0.2$ for horizontal growth ($\Gamma<1$). To reproduce different snowflake morphologies, parameters $\Gamma$, $u_0$, $L_{\text{sat}}$ and $\epsilon_{z}$  were varied. All study case parameters  are gathered in Table \ref{tab1}. The simulation method is provided in appendix \ref{annexe2} and \ref{annexe3}.


\newcounter{foo}
\begin{table}[ht]
\centering
\begin{tabular}{l c c c c c}
   \hline
  & $\Gamma$  & $u_0$ & $L_{\text{sat}}$  & $\epsilon_{z}$  \\  
   \hline
\footnotesize\refstepcounter{foo}\thefoo\label{stellard} stellar dendrite &  0.5& 0.7& 1.0& 0.05    \\ 
\footnotesize\refstepcounter{foo}\thefoo\label{fernd}  fern dendrite      &    0.5& 0.8& 1.6& 0.05    \\ 
\footnotesize\refstepcounter{foo}\thefoo\label{plated}  dendritic arms plate            &    0.4& 0.6& 1.0& 0.1    \\ 
\footnotesize\refstepcounter{foo}\thefoo\label{stellarp}    stellar plate            &    0.5& 0.5& 1.8& 0.3    \\  
\footnotesize\refstepcounter{foo}\thefoo\label{star}  \& \footnotesize\refstepcounter{foo}\thefoo\label{12star}  stars   &   0.5& 0.5& 1.0& 0.3     \\ 
\footnotesize\refstepcounter{foo}\thefoo\label{sandwich}  double plate             &   0.4& 0.5& 1.6& 0.1    \\ 
\footnotesize\refstepcounter{foo}\thefoo\label{sectoredA}   sectored plate &   0.4& 0.5(0.8)& 1.0& 0.25(0.1)\\
\footnotesize\refstepcounter{foo}\thefoo\label{solid}  solid plate             &  0.25& 0.4& 2.0& 0.2    \\ 
\footnotesize\refstepcounter{foo}\thefoo\label{scrolls}  scrolls on plate    &    0.4(0.2)& 0.5(0.8)& 1.0& 0.25(0.4)    \\
\footnotesize\refstepcounter{foo}\thefoo\label{needle} \& \footnotesize\refstepcounter{foo}\thefoo\label{needlec}  needles             &   3.0& 0.8& 1.0& 0.5    \\ 
\footnotesize\refstepcounter{foo}\thefoo\label{prism}  hollow prism            &  3.0& 0.3& 3.0& 0.5     \\ 
\footnotesize\refstepcounter{foo}\thefoo\label{cappedA}-\footnotesize\refstepcounter{foo}\thefoo\label{cappedB}  capped column &   5.0(0.2)& 0.8& 1.0(2.0/1.0)& 0.5(0.4) \\ 
   \hline
\end{tabular}
\caption{Parameters used in the simulation of different morphologies of snowflakes. Parenthesis correspond to the case where parameter were changed during simulations.}
\label{tab1}
\end{table}



\begin{figure}[ht]
\centering
\subfigure[~~Fern dendrite \ref{fernd}]{\includegraphics[height=3.9cm]{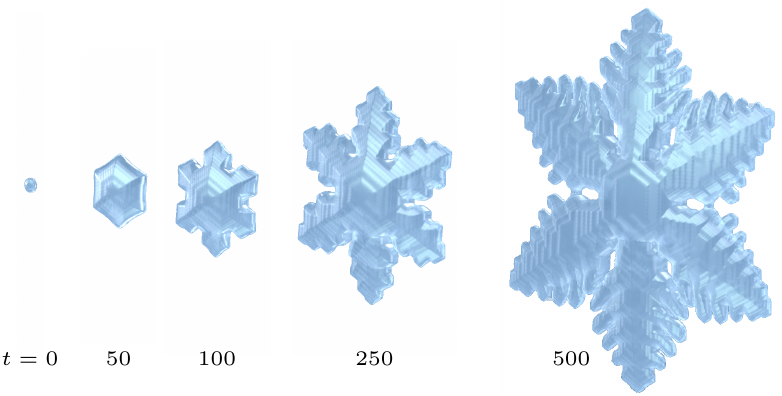}\label{fig:fern_ev}}\vspace{-0.6cm}
\subfigure[~~Hollow prism \ref{prism}]{\includegraphics[height=3.3cm]{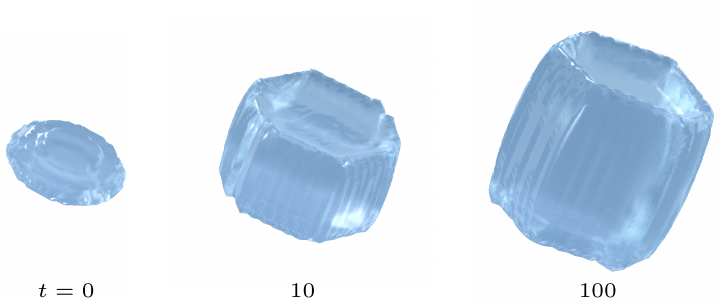}\label{fig:prism_ev}}
\caption{Isosurface representation ($\phi=0$) of snowflake growth, using the software Blender for visual rendering. Top: fern dendrite \ref{fernd}. The growth mechanism is essentially two dimensional \cite{libbrecht2005physics}. First, the initial disk grows into a transient flat faceted hexagon. Then, the branching instability  occurs in the horizontal plane ($t=50$), for a critical crystal radius. The resulting faceted dendrites are equipped with developed faceted side branches. Bottom: prism \ref{prism}. The seed grows vertically into a prism aligned on the Wulff shape. Then the basal facet breaks, and a hollow is formed at $t=10$.  Time is indicated  in $\tau_0$ unit.}
\label{fig:evolution} 
\end{figure}

Two limit cases of snowflake growth are displayed in \reffig{fig:evolution}. \reffig{fig:fern_ev} shows the different stages of the growth of a fernlike dendrite \ref{fernd} (p. 59 of \cite{libbrecht2006ken}), and \reffig{fig:prism_ev} details the formation of a hollow prism \ref{prism} (p 64-66 of \cite{libbrecht2006ken}).

For both cases, the growth stages are very similar to the experimental growth kinetics obtained by Libbrecht \cite{siteLibbrecht}: the snowflake first aligns on the equilibrium Wulff shape, until the Mullins-Sekerka instability \cite{mullins1964stability} occurs for a critical crystal radius. This instability is related to the Berg effect \cite{berg1938crystal}, stating that the supersaturation field around a faceted snow crystal is largest at facet edges. Instability thus occurs when the snowflake reaches a critical radius \cite{fujioka1974morphological}, and kinetic effects at corners overcome the highly anisotropic surface tension \cite{libbrecht2005physics}.

In \reffig{fig:fern_ev}, the simulated fernlike dendrite morphology differs from the non axisymmetric shape predicted by the theory of Brener \cite{brener1996three}, usually used to describe dendrites with cubic symmetry \cite{singer2004three,cartalade2016lattice}. We suggest this is due to the strong vertical anisotropy flattening, and to faceting, which limits the formation of a horizontal lacuna at the tail of dendrites. The obtained feathery shape with a transversally sharp tip rather resembles Furukawa's experimental ice crystals \cite{furukawa1993three}. Besides, experimental snowflakes display characteristic surface patterns, such as ridges \cite{gravner2009modeling}, and flat basal planes forming steps \cite{furukawa1993three}. Such patterns are also reproduced by our model in \reffig{fig:fern_ev}. At low temperatures, surface nucleation and spiral growth are the leading growth processes on the basal faces for horizontal growth \cite{nelson2001growth}. Though such mechanisms are not explicitly included in our model, the presence of surface steps in our simulations evokes the terrace growth resulting from nucleation, experimentally observed on ice crystals \cite{sazaki2010elementary}. It can be underlined that the succession of flat basal planes on both real and simulated snowflakes, reflects the main stages of snowflake history. For instance, the hexagonal Wullf shape before branching instability is clearly memorized in \reffig{fig:fern_ev}. This is consistent with real snowflakes observation, provided in appendix \ref{annexe1}.


\reffig{fig:RV} displays the time evolution of snowflake size $L$, defined as the arm length for horizontal growth (blue), and the column length for vertical growth (red).

\begin{figure}[ht]
\centering
\includegraphics{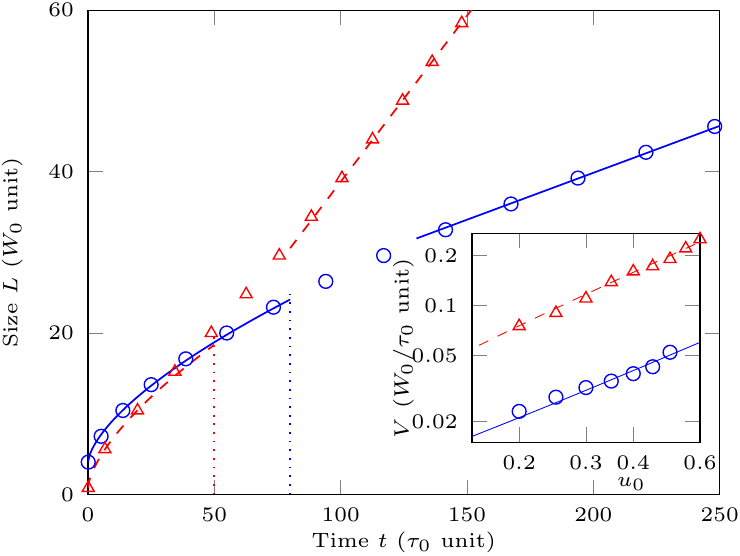}
\caption{Snowflake size $L$ ($W_0$ units) vs. time $t$ ($\tau_0$ units) (marks), for $u_0=0.6$. Two growth regimes appear, with a transition at $t=50$ for vertical growth, and $t=80$ for horizontal growth (dots). First regime: $L(t)\propto t^{\alpha}$ ($\rho=0.99$), with $\alpha=0.62$ for horizontal growth (line), and $\alpha=0.63$ for vertical growth (dashes). Second regime: $L(t)\propto A t$ ($\rho=0.99$), with $A=0.12$ for horizontal growth (line), and $A=0.41$ for vertical growth (dashes).  Insert: tip velocity $V$ ($W_0/\tau_0$ units) vs. reduced supersaturation $u_0$ (marks). $V=A' u_0$ ($\rho=0.99$) with $A'=0.62$ for horizontal growth (line), and $A'=0.62$ for vertical growth (dashes). Blue: horizontal growth (parameters \ref{fernd} except $u_0$ varied from 0.2 to 0.6). Red: vertical growth (parameters \ref{needle} except $u_0$ varied from 0.2 to 0.6).}
\label{fig:RV} 
\end{figure} 

It can be seen that two growth regimes appear. After a first growth regime, when both dynamics are similar, the needle growth accelerates, while plate-like growth slows down. In the case of horizontal growth, Libbrecht experimentally noted in \cite{libbrecht2016toward}, that this change of regime coincides with full faceting occurrence on the prismatic face, causing growth to decelerate. In the case of vertical growth, it was also observed in \cite{libbrecht2016toward}, that the growth acceleration was related to the formation of a vicinal surface at the needle tip, which enhances water attachment at the tip. During the first regime, the crystal growth is slower than diffusion, and the system satisfies the Laplacian approximation. Within this framework, Algrem et al. found in \cite{almgren1993scaling}, that arm growth should display the self similar scaling behaviour $L(t)\propto t^{3/5}$. This is confirmed by our simulations. It is interesting to note that the disparity of our simulations with the $t^{1/2}$ diffusive law of  Zener et al. for spherical precipitates \cite{zener1949theory}, is due to surface tension.

In the second regime, the growth kinetics is linear (line and dashes in \reffig{fig:RV}), consistently with the selection theory \cite{amar1993theory,saito1996statistical}. The associated tip velocity $V$ is thus the slope of this linear function. Using the same procedure for different $u_0$, $V$ versus $u_0$ could be plotted in the insert. It appears that $V$ is proportional to $u_0$. Therefore, the growth velocity of both faceted and non faceted dendrites satisfies the universal law \cite{langer1978evidence}.

\begin{figure}[H]
\centering
\begin{tabular}{ccc}
   \multicolumn{2}{c}{$\Gamma< 1$}  &  $\Gamma\geq 1$ \\
\includegraphics[height=2.5cm]{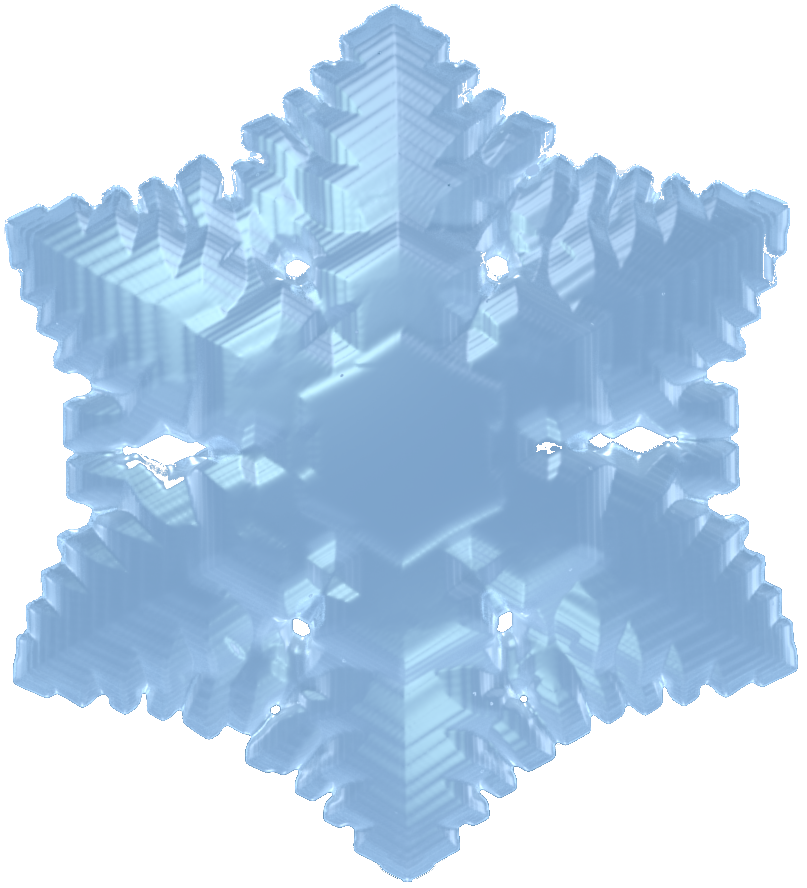} & \includegraphics[height=2.5cm]{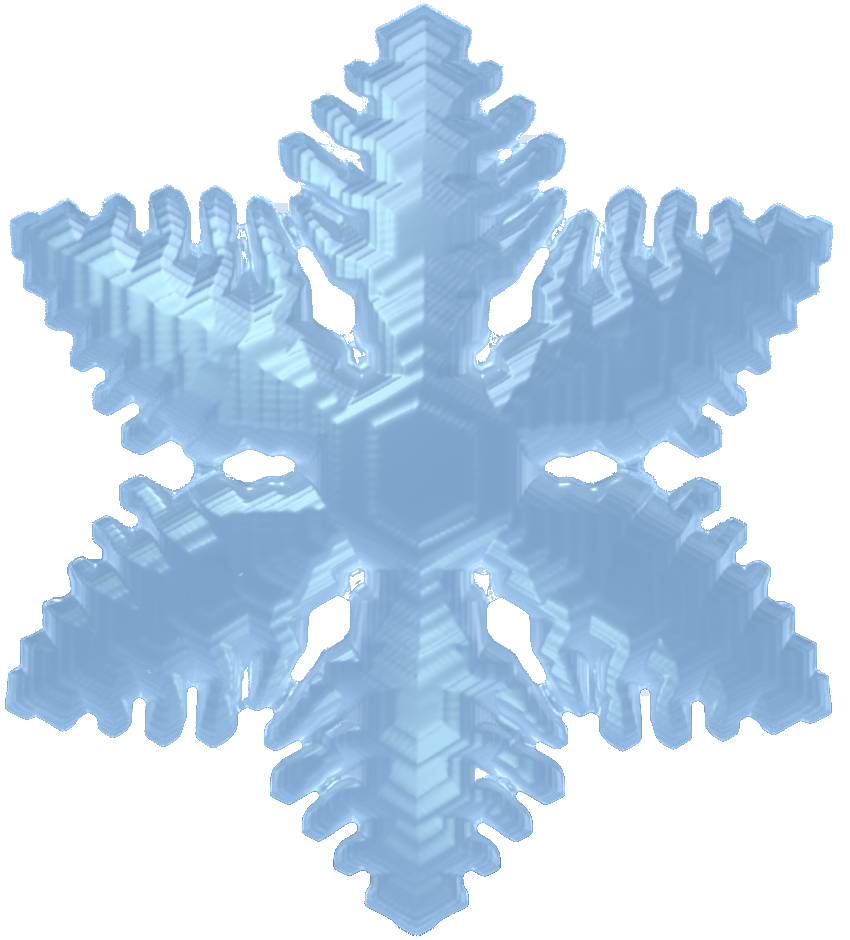} & \includegraphics[height=2.5cm]{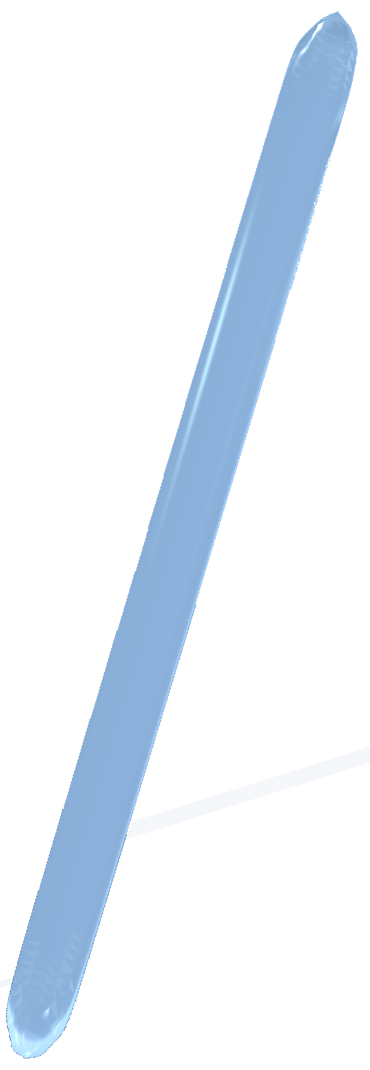} \\

 \scriptsize \ref{stellard} - stellar dendrite & \scriptsize\ref{fernd} - fern dendrite & \scriptsize\ref{needle} - needle\\

   \includegraphics[height=2.5cm]{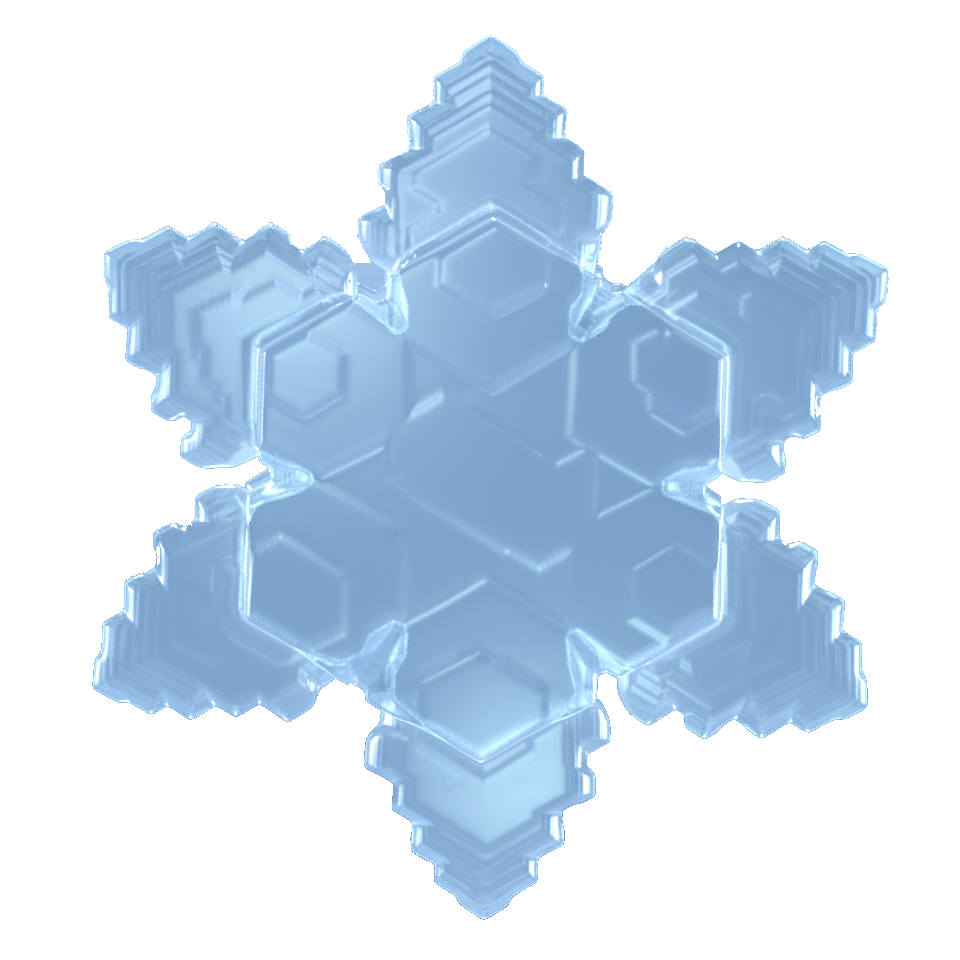} & \includegraphics[height=2.5cm]{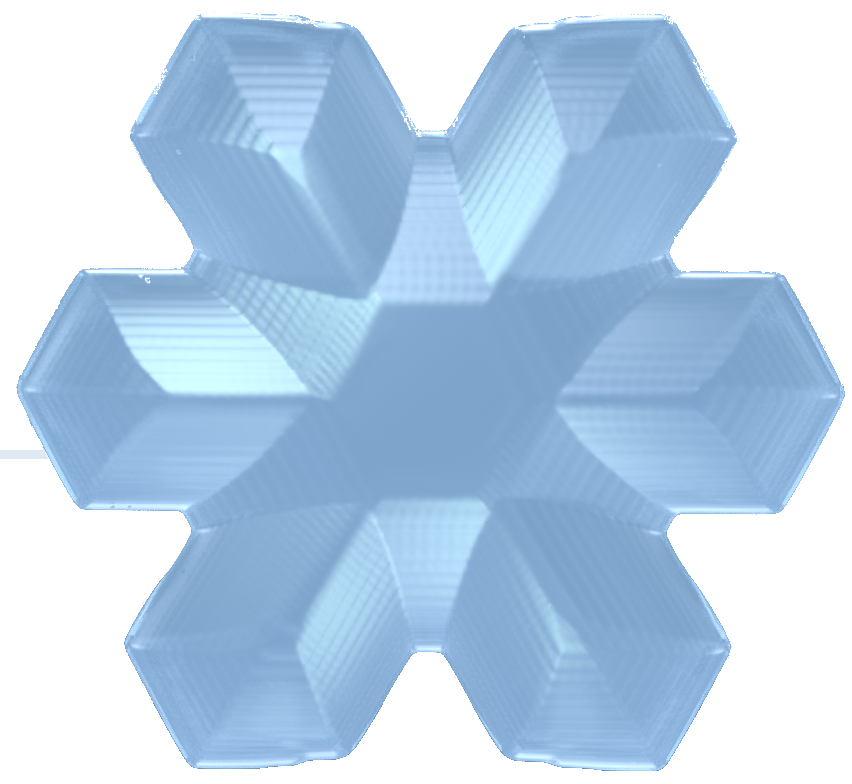} & \includegraphics[height=2.5cm]{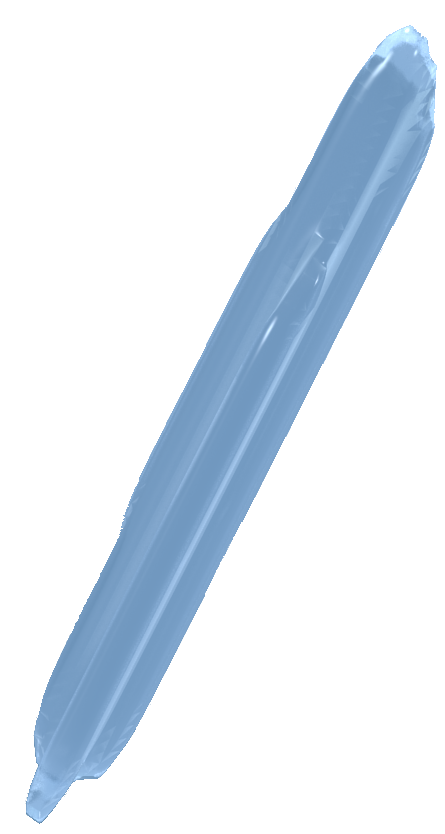} \\

  \scriptsize\ref{plated} - dendrite plate  &\scriptsize\ref{stellarp} - stellar plate  & \scriptsize\ref{needlec} needles cluster \\
   \includegraphics[height=2.5cm]{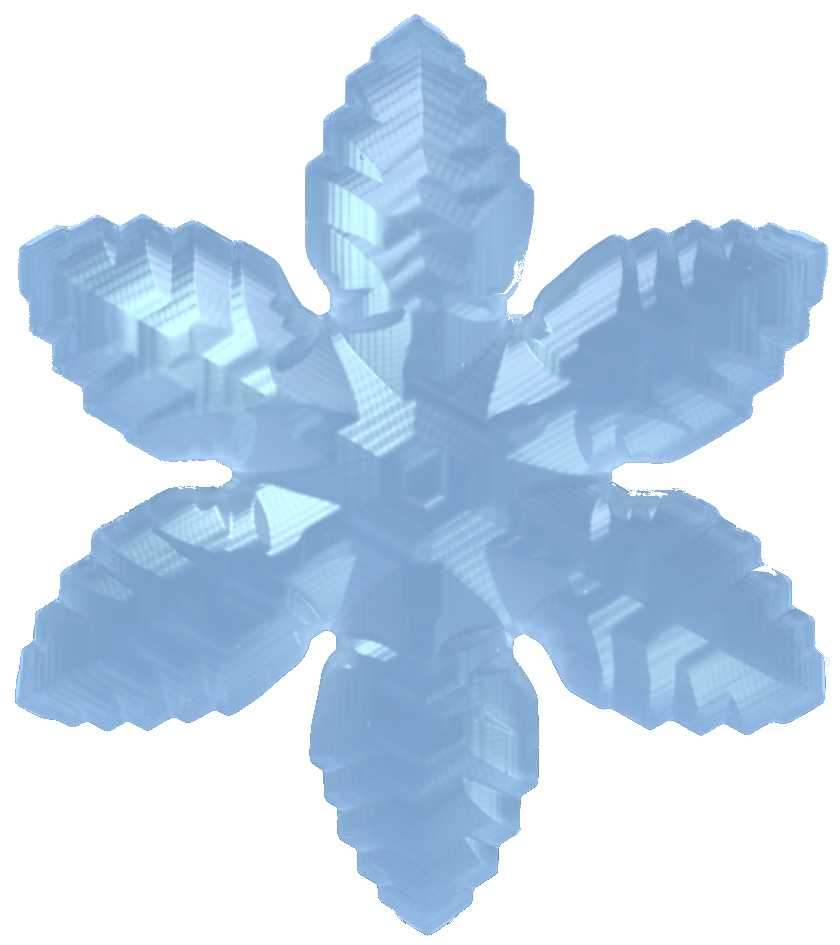} & \includegraphics[height=2.5cm]{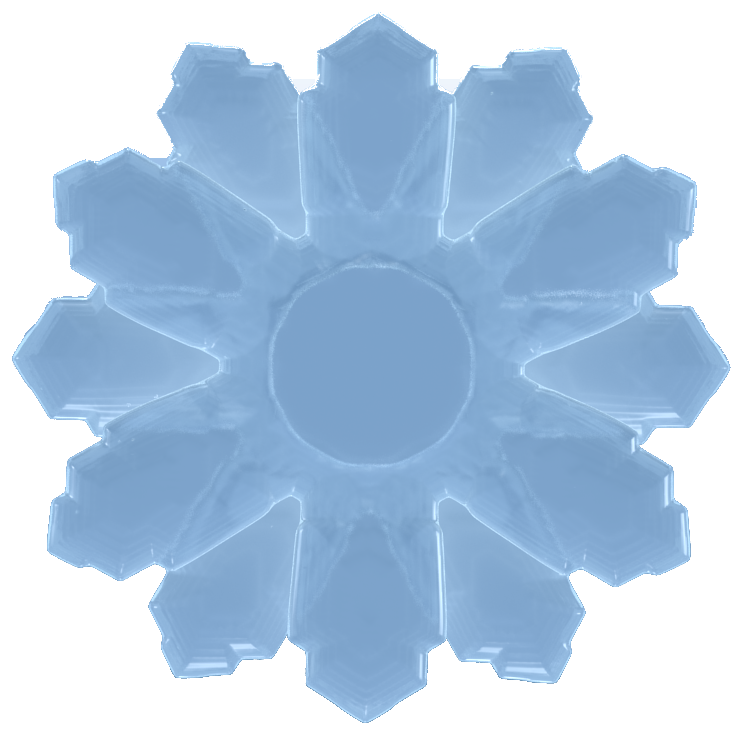} & \includegraphics[height=2.5cm]{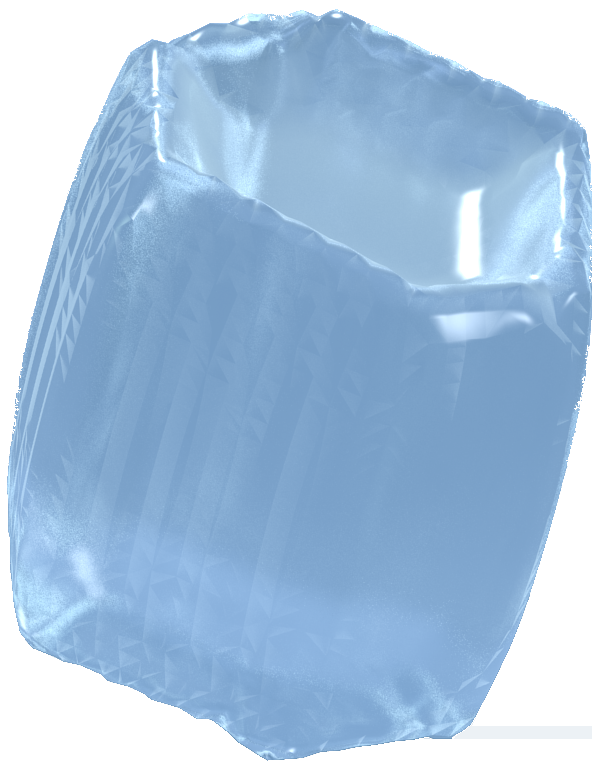}\\

    \scriptsize\ref{star} - simple star& \scriptsize\ref{12star} - 12-star & \scriptsize\ref{prism} - hollow prism \\
 
  \includegraphics[height=2.5cm]{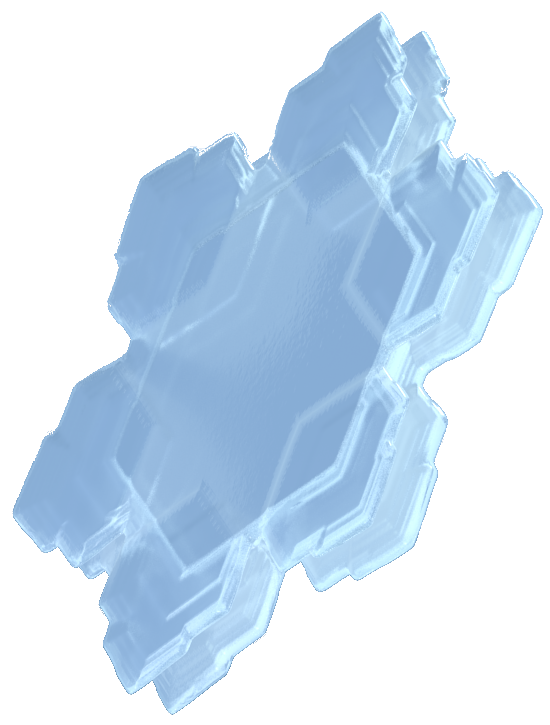}  &  \includegraphics[height=2.5cm]{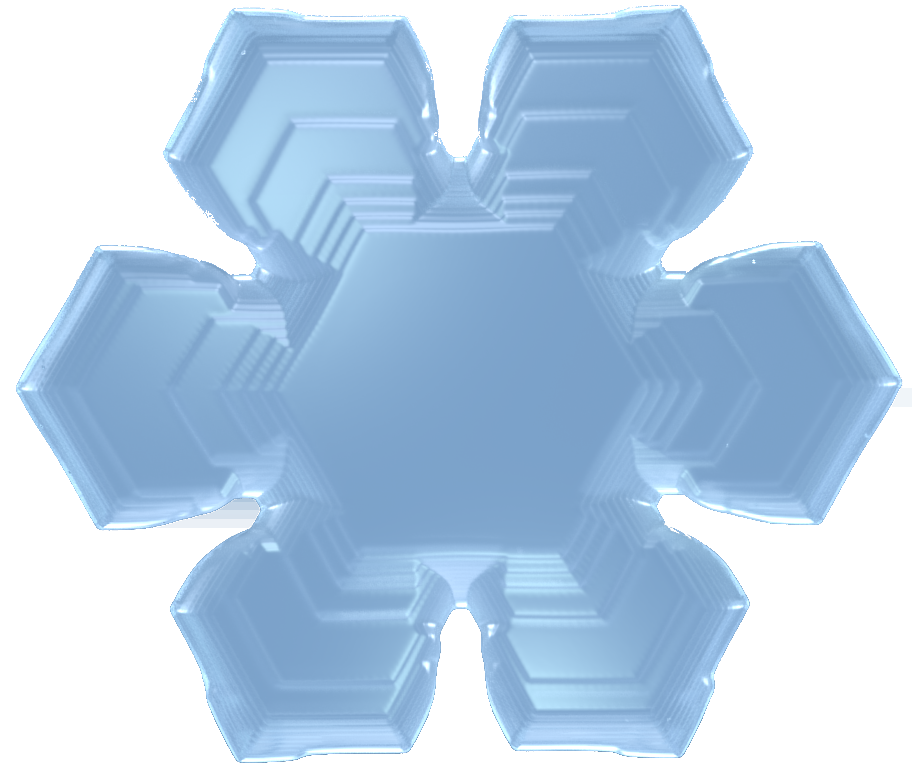}  &  \includegraphics[height=2.2cm]{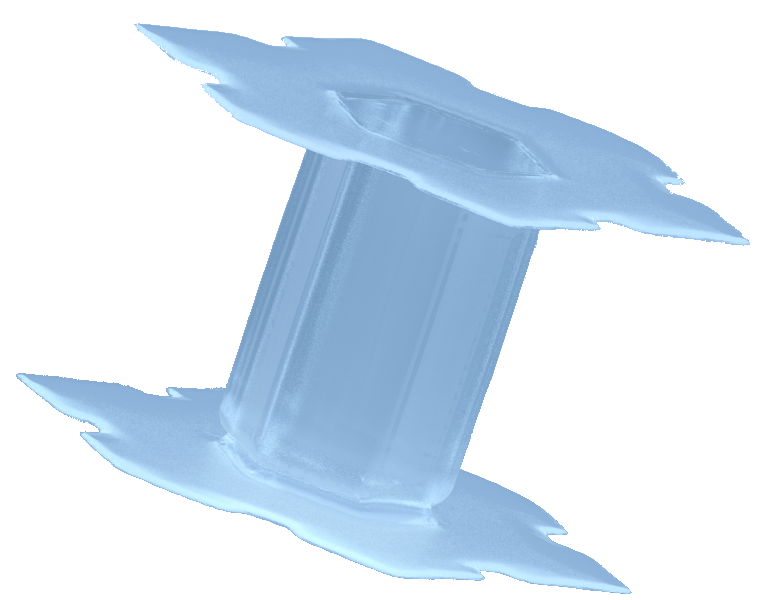} \\

  \scriptsize\ref{sandwich} - double plate  & \scriptsize\ref{sectoredA} - sectored plate & \scriptsize\ref{cappedA} - capped col. a \\
   \includegraphics[height=2.2cm]{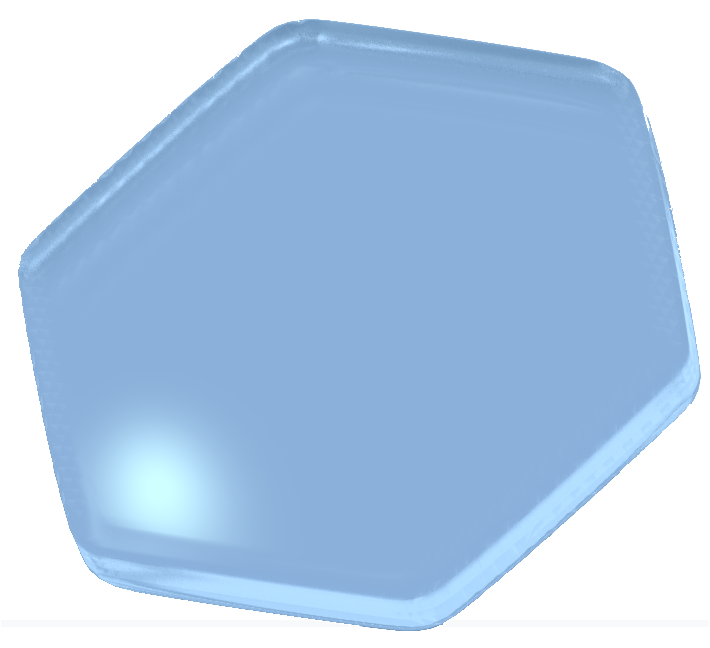}  &  \includegraphics[height=2.5cm]{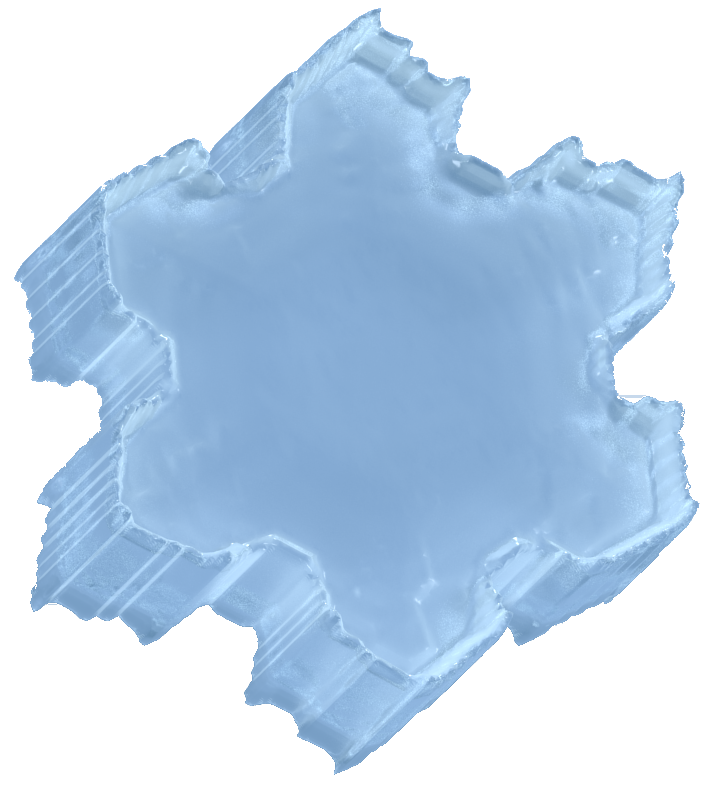} &  \includegraphics[height=2.5cm]{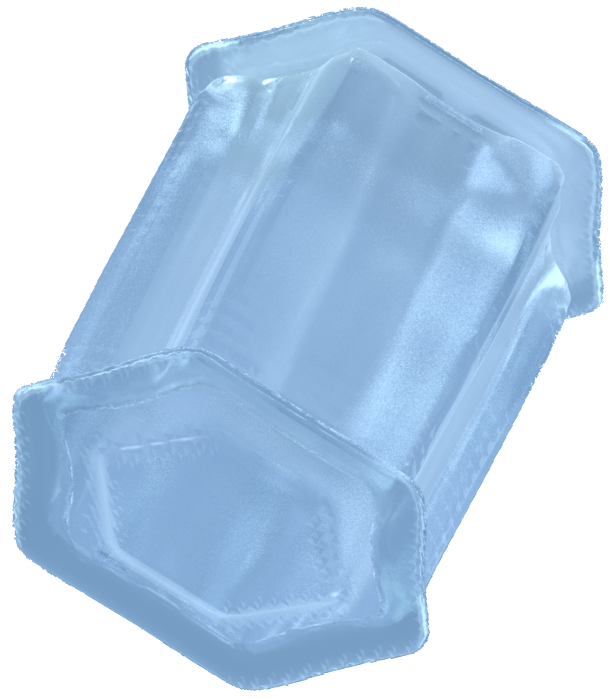} \\

  \scriptsize\ref{solid} - solid plate & \scriptsize\ref{scrolls} - scrolls on plate & \scriptsize\ref{cappedB} - capped col. b\\
\end{tabular}
\caption{Simulated snowflake classification, using the software Blender for visual rendering. $\Gamma<1$ gives the flat snow crystals \ref{stellard} to  \ref{solid}. $\Gamma\geq 1$ gives columns \ref{needle} \& \ref{prism}. Primary habit switch is achieved by shifting $\Gamma$ during simulations. $\Gamma$ from 3.0 to 0.2 at $t=70$, gives capped columns \ref{cappedA} and \ref{cappedB}. $\Gamma$ from 0.4 to 3.0 gives scrolls on plate \ref{scrolls}. For $\Gamma<1$, no facet breaking happens when $u_0\leq 0.4$ (solid  plate \ref{solid}). Branching instability appears for $u_0=0.5$ (snowflakes \ref{plated} to \ref{sectoredA}), and side branching requires  $u_0\geq 0.7$ (dendrites \ref{stellard} and \ref{fernd}). For $\Gamma>1$, $u_0=0.3$ gives a prism \ref{prism}, and $u_0=0.8$ induces vertical branching with needles \ref{needle} and \ref{needlec}. $L_{\text{sat}}=1.0$ gives a thick side branching for the  stellar dendrite \ref{stellard}. Rising $L_{\text{sat}}$ to $L_{\text{sat}}=1.6$ refines side branches with the fern dendrite \ref{stellard}. Higher values of $L_{\text{sat}}$ destroy side branching, and lead to branches maximum faceting with the stellar plate \ref{stellarp} ($L_{\text{sat}}=1.8$).}
\label{fig:table}
\end{figure}

It is consistent with previous experimental \cite{libbrecht2002electrically}, and numerical observations \cite{debierre2003phase,barrett2012numerical}. Langer et al. argued in \cite{langer1978evidence}, that snowflake dendrites frontmost tip is molecularly rough for both non-faceted and faceted dendrites. The attachment kinetics can thus be considered isotropic with circular symmetry.

\reffig{fig:table} compiles equilibrium shapes of simulated snowflakes, using the parameters in table \ref{tab1}. An achievement of our model is to recover the principal snowflake morphologies of the Nakaya diagram \cite{nakaya1951formation}, depending exclusively on four parameters: $\Gamma$, $u_0$, $L_{\text{sat}}$ and $\epsilon_z$.

The first parameter $\Gamma$ fully determines the primary habit in our model. We suggest this parameter may be fitted on the yield between prismatic and basal attachment coefficients \cite{libbrecht2008crystal}. This yield accounts for the alternation of dominating growth mechanisms with the temperature. Two theories are proposed to explain this alternation of growth preference: surface nucleation for low temperatures \cite{nelson2001growth}, and quasi-liquid layer growth causing surface roughening \cite{kuroda1982growth} near the melting point \cite{libbrecht2005physics}. Then, Hertz-Knudsen relation for facet normal velocity \cite{saito1996statistical} makes supersaturation $u_0$ the priming parameter in facet instability and growth velocity \cite{nelson2001growth}. In the phase field model, the impact of $u_0$ is reinterpreted as the thermodynamic driving force, and higher values also favour branching instability and faster growth \cite{singer2008phase}. The parameter $L_{\text{sat}}$ controls the compactness of branching for non faceted growth \cite{kobayashi1993modeling}. Here, low values foster branching instability, while raising $L_{\text{sat}}$ drives the system toward quasistatic diffusion, and fosters a stronger faceting. As for the parameter $\epsilon_z$, it mainly influences the formation of surface patterns for horizontal growth. 

Especially interesting are two model examples. First, the double plate \ref{sandwich} (p. 75 in \cite{libbrecht2006ken}), results from the \emph{sandwich instability} of a flat snowflake, and the twelve arms star \ref{12star} was simulated by the aggregation of two simple stars, with a $30^{\circ}$ tilt \cite{Kikuchi1998169}.

Finally, simulated snowflake morphology depends on temperature and density excess over vapour/water equilibrium  $\rho_{\text{super}}^W=\rho-\rho_{\text{sat}}^W$  (g.m$^{-3}$). It is quantitatively consistent with the Nakaya diagram (see \cite{marshall1954theory}). Using $u_0=(\rho_{\text{super}}^W+\rho_{\text{sat}}^W-\rho_{\text{sat}}^I)/\rho_{\text{sat}}^I$, where $\rho_{\text{sat}}^I$ is the vapour/ice equilibrium density, and fitting $\Gamma$ on \cite{libbrecht2008crystal}, our simulations predict plate formation near $(T,\rho_{\text{super}}^W)=(-19,0.15)$ and $(-12,0.4)$, versus $(-19,0.15)$, $(-12,0.4)$, but also  $(-10,0.18)$ in the Nakaya diagram. Simulated dendrites occur near $(-16,0.4)$, versus $(-15,0.4)$ and $(-15,0.25)$ for experiments.  Finally, hollow columns are formed at $(-8,0.6)$ in our simulations, versus $(-8,0.6)$ and $(-8,0.3)$ in the Nakaya diagram. However, small supersaturations $\rho_{\text{sat}}^I<\rho<\rho_{\text{sat}}^W$ are beyond reach for our model. This shortcoming is complementary to the model of Barrett et al. \cite{barrett2012numerical}. Indeed, the latter is bound to $\rho\leq \rho_{\text{sat}}^W$ or $T\to 0^{\circ}$C, due to the Laplacian approximation for small supersaturations \cite{libbrecht2005physics}. Our model on the contrary requires larger values for $L_{\text{sat}}$, and greater computational resources are needed to reach such supersaturations.

In this paper, we have shown that the proposed modified phase field model is able to reproduce the complex dynamics of snowflake growth. Benefiting from the modularity and versatility of the phase field model, this approach can be extended to describe the growth of snowflakes in the real atmospheric conditions. It can notably be equipped with stochastic effects, for the asymmetric growth of more exotic morphologies of snowflakes in the Nakaya diagram. Continuum fluid dynamics can also be easily introduced in the kinetics equation \refeq{1} and \refeq{2}, to simulate the air flow around the snowflake during its fall from clouds \cite{PhysRevE.64.041602}. Besides, this approach is not restricted to snowflake growth. The developed approach opens a new way to answer numerous outstanding questions concerning dendritic growth in wide range of materials. We can cite, for instance, the faceted dendritic growth observed in pure isotactic polystyrene films, or the snowflake-like growth of graphene monocrystal \cite{whiteway2015time}. 

\section*{Acknowledgements}

This work was supported in part by the grant from LABEX "Cistic". The simulations were performed at the Centre de Ressources Informatiques de Haute-Normandie (CRIHAN) and at the IDRIS of CNRS.

\section*{Author Contributions}

G.D. and H.Z. developed the model and wrote the manuscript. R.P. participated in the code development. All authors analyzed the data, discussed the results and commented on the manuscript.

\bibliographystyle{apsrev4-1}


\begin{thebibliography}{45}%
\makeatletter
\providecommand \@ifxundefined [1]{%
 \@ifx{#1\undefined}
}%
\providecommand \@ifnum [1]{%
 \ifnum #1\expandafter \@firstoftwo
 \else \expandafter \@secondoftwo
 \fi
}%
\providecommand \@ifx [1]{%
 \ifx #1\expandafter \@firstoftwo
 \else \expandafter \@secondoftwo
 \fi
}%
\providecommand \natexlab [1]{#1}%
\providecommand \enquote  [1]{``#1''}%
\providecommand \bibnamefont  [1]{#1}%
\providecommand \bibfnamefont [1]{#1}%
\providecommand \citenamefont [1]{#1}%
\providecommand \href@noop [0]{\@secondoftwo}%
\providecommand \href [0]{\begingroup \@sanitize@url \@href}%
\providecommand \@href[1]{\@@startlink{#1}\@@href}%
\providecommand \@@href[1]{\endgroup#1\@@endlink}%
\providecommand \@sanitize@url [0]{\catcode `\\12\catcode `\$12\catcode
  `\&12\catcode `\#12\catcode `\^12\catcode `\_12\catcode `\%12\relax}%
\providecommand \@@startlink[1]{}%
\providecommand \@@endlink[0]{}%
\providecommand \url  [0]{\begingroup\@sanitize@url \@url }%
\providecommand \@url [1]{\endgroup\@href {#1}{\urlprefix }}%
\providecommand \urlprefix  [0]{URL }%
\providecommand \Eprint [0]{\href }%
\providecommand \doibase [0]{http://dx.doi.org/}%
\providecommand \selectlanguage [0]{\@gobble}%
\providecommand \bibinfo  [0]{\@secondoftwo}%
\providecommand \bibfield  [0]{\@secondoftwo}%
\providecommand \translation [1]{[#1]}%
\providecommand \BibitemOpen [0]{}%
\providecommand \bibitemStop [0]{}%
\providecommand \bibitemNoStop [0]{.\EOS\space}%
\providecommand \EOS [0]{\spacefactor3000\relax}%
\providecommand \BibitemShut  [1]{\csname bibitem#1\endcsname}%
\let\auto@bib@innerbib\@empty
\bibitem [{\citenamefont {Pamuk}\ \emph {et~al.}(2012)\citenamefont {Pamuk},
  \citenamefont {Soler}, \citenamefont {Ram\'{\i}rez}, \citenamefont {Herrero},
  \citenamefont {Stephens}, \citenamefont {Allen},\ and\ \citenamefont
  {Fern\'andez-Serra}}]{PhysRevLett.108.193003}%
  \BibitemOpen
  \bibfield  {author} {\bibinfo {author} {\bibfnamefont {B.}~\bibnamefont
  {Pamuk}}, \bibinfo {author} {\bibfnamefont {J.~M.}\ \bibnamefont {Soler}},
  \bibinfo {author} {\bibfnamefont {R.}~\bibnamefont {Ram\'{\i}rez}}, \bibinfo
  {author} {\bibfnamefont {C.~P.}\ \bibnamefont {Herrero}}, \bibinfo {author}
  {\bibfnamefont {P.~W.}\ \bibnamefont {Stephens}}, \bibinfo {author}
  {\bibfnamefont {P.~B.}\ \bibnamefont {Allen}}, \ and\ \bibinfo {author}
  {\bibfnamefont {M.-V.}\ \bibnamefont {Fern\'andez-Serra}},\ }\href@noop {}
  {\bibfield  {journal} {\bibinfo  {journal} {Phys. Rev. Lett.}\ }\textbf
  {\bibinfo {volume} {108}},\ \bibinfo {pages} {193003} (\bibinfo {year}
  {2012})}\BibitemShut {NoStop}%
\bibitem [{\citenamefont {Ball}(2016)}]{ball2016material}%
  \BibitemOpen
  \bibfield  {author} {\bibinfo {author} {\bibfnamefont {P.}~\bibnamefont
  {Ball}},\ }\href@noop {} {\bibfield  {journal} {\bibinfo  {journal} {Nature
  Materials}\ }\textbf {\bibinfo {volume} {15}},\ \bibinfo {pages} {1060}
  (\bibinfo {year} {2016})}\BibitemShut {NoStop}%
\bibitem [{\citenamefont {Nakaya}\ \emph {et~al.}(1938)\citenamefont {Nakaya},
  \citenamefont {Sat{\^o}},\ and\ \citenamefont
  {Sekido}}]{nakaya1938preliminary}%
  \BibitemOpen
  \bibfield  {author} {\bibinfo {author} {\bibfnamefont {U.}~\bibnamefont
  {Nakaya}}, \bibinfo {author} {\bibfnamefont {I.}~\bibnamefont {Sat{\^o}}}, \
  and\ \bibinfo {author} {\bibfnamefont {Y.}~\bibnamefont {Sekido}},\
  }\href@noop {} {\bibfield  {journal} {\bibinfo  {journal} {Journal of the
  Faculty of Science, Hokkaido Imperial University. Ser. 2}\ }\textbf {\bibinfo
  {volume} {2}},\ \bibinfo {pages} {1} (\bibinfo {year} {1938})}\BibitemShut
  {NoStop}%
\bibitem [{\citenamefont {Magono}\ \emph {et~al.}(1966)\citenamefont {Magono},
  \citenamefont {Chung} \emph {et~al.}}]{magono1966meteorological}%
  \BibitemOpen
  \bibfield  {author} {\bibinfo {author} {\bibfnamefont {C.}~\bibnamefont
  {Magono}}, \bibinfo {author} {\bibfnamefont {W.}~\bibnamefont {Chung}},
  \emph {et~al.},\ }\href@noop {} {\bibfield  {journal} {\bibinfo  {journal}
  {Journal of the Faculty of Science, Hokkaido University. Series 7,
  Geophysics}\ }\textbf {\bibinfo {volume} {2}},\ \bibinfo {pages} {321}
  (\bibinfo {year} {1966})}\BibitemShut {NoStop}%
\bibitem [{\citenamefont {Nakaya}(1951)}]{nakaya1951formation}%
  \BibitemOpen
  \bibfield  {author} {\bibinfo {author} {\bibfnamefont {U.}~\bibnamefont
  {Nakaya}},\ }\href@noop {} {\bibfield  {journal} {\bibinfo  {journal}
  {Compendium Meteor}\ ,\ \bibinfo {pages} {207}} (\bibinfo {year}
  {1951})}\BibitemShut {NoStop}%
\bibitem [{\citenamefont {Libbrecht}(2005)}]{libbrecht2005physics}%
  \BibitemOpen
  \bibfield  {author} {\bibinfo {author} {\bibfnamefont {K.~G.}\ \bibnamefont
  {Libbrecht}},\ }\href@noop {} {\bibfield  {journal} {\bibinfo  {journal}
  {Reports on progress in physics}\ }\textbf {\bibinfo {volume} {68}},\
  \bibinfo {pages} {855} (\bibinfo {year} {2005})}\BibitemShut {NoStop}%
\bibitem [{\citenamefont {Mason}(1992)}]{mason1992snow}%
  \BibitemOpen
  \bibfield  {author} {\bibinfo {author} {\bibfnamefont {B.~J.}\ \bibnamefont
  {Mason}},\ }\href@noop {} {\bibfield  {journal} {\bibinfo  {journal}
  {Contemporary physics}\ }\textbf {\bibinfo {volume} {33}},\ \bibinfo {pages}
  {227} (\bibinfo {year} {1992})}\BibitemShut {NoStop}%
\bibitem [{\citenamefont {Kuroda}\ and\ \citenamefont
  {Lacmann}(1982)}]{kuroda1982growth}%
  \BibitemOpen
  \bibfield  {author} {\bibinfo {author} {\bibfnamefont {T.}~\bibnamefont
  {Kuroda}}\ and\ \bibinfo {author} {\bibfnamefont {R.}~\bibnamefont
  {Lacmann}},\ }\href@noop {} {\bibfield  {journal} {\bibinfo  {journal}
  {Journal of Crystal Growth}\ }\textbf {\bibinfo {volume} {56}},\ \bibinfo
  {pages} {189} (\bibinfo {year} {1982})}\BibitemShut {NoStop}%
\bibitem [{\citenamefont {Nelson}(2001)}]{nelson2001growth}%
  \BibitemOpen
  \bibfield  {author} {\bibinfo {author} {\bibfnamefont {J.}~\bibnamefont
  {Nelson}},\ }\href@noop {} {\bibfield  {journal} {\bibinfo  {journal}
  {Philosophical Magazine A}\ }\textbf {\bibinfo {volume} {81}},\ \bibinfo
  {pages} {2337} (\bibinfo {year} {2001})}\BibitemShut {NoStop}%
\bibitem [{\citenamefont {Nada}\ and\ \citenamefont
  {Furukawa}(1996)}]{nada1996anisotropic}%
  \BibitemOpen
  \bibfield  {author} {\bibinfo {author} {\bibfnamefont {H.}~\bibnamefont
  {Nada}}\ and\ \bibinfo {author} {\bibfnamefont {Y.}~\bibnamefont
  {Furukawa}},\ }\href@noop {} {\bibfield  {journal} {\bibinfo  {journal}
  {Journal of crystal growth}\ }\textbf {\bibinfo {volume} {169}},\ \bibinfo
  {pages} {587} (\bibinfo {year} {1996})}\BibitemShut {NoStop}%
\bibitem [{\citenamefont {Furukawa}\ and\ \citenamefont
  {Nada}(1997)}]{furukawa1997anisotropic}%
  \BibitemOpen
  \bibfield  {author} {\bibinfo {author} {\bibfnamefont {Y.}~\bibnamefont
  {Furukawa}}\ and\ \bibinfo {author} {\bibfnamefont {H.}~\bibnamefont
  {Nada}},\ }\href@noop {} {\bibfield  {journal} {\bibinfo  {journal} {The
  Journal of Physical Chemistry B}\ }\textbf {\bibinfo {volume} {101}},\
  \bibinfo {pages} {6167} (\bibinfo {year} {1997})}\BibitemShut {NoStop}%
\bibitem [{\citenamefont {Benet}\ \emph {et~al.}(2016)\citenamefont {Benet},
  \citenamefont {Llombart}, \citenamefont {Sanz},\ and\ \citenamefont
  {MacDowell}}]{benet2016premelting}%
  \BibitemOpen
  \bibfield  {author} {\bibinfo {author} {\bibfnamefont {J.}~\bibnamefont
  {Benet}}, \bibinfo {author} {\bibfnamefont {P.}~\bibnamefont {Llombart}},
  \bibinfo {author} {\bibfnamefont {E.}~\bibnamefont {Sanz}}, \ and\ \bibinfo
  {author} {\bibfnamefont {L.~G.}\ \bibnamefont {MacDowell}},\ }\href@noop {}
  {\bibfield  {journal} {\bibinfo  {journal} {Physical Review Letters}\
  }\textbf {\bibinfo {volume} {117}},\ \bibinfo {pages} {096101} (\bibinfo
  {year} {2016})}\BibitemShut {NoStop}%
\bibitem [{\citenamefont {Gravner}\ and\ \citenamefont
  {Griffeath}(2009)}]{gravner2009modeling}%
  \BibitemOpen
  \bibfield  {author} {\bibinfo {author} {\bibfnamefont {J.}~\bibnamefont
  {Gravner}}\ and\ \bibinfo {author} {\bibfnamefont {D.}~\bibnamefont
  {Griffeath}},\ }\href@noop {} {\bibfield  {journal} {\bibinfo  {journal}
  {Physical Review E}\ }\textbf {\bibinfo {volume} {79}},\ \bibinfo {pages}
  {011601} (\bibinfo {year} {2009})}\BibitemShut {NoStop}%
\bibitem [{\citenamefont {Barrett}\ \emph {et~al.}(2012)\citenamefont
  {Barrett}, \citenamefont {Garcke},\ and\ \citenamefont
  {N{\"u}rnberg}}]{barrett2012numerical}%
  \BibitemOpen
  \bibfield  {author} {\bibinfo {author} {\bibfnamefont {J.~W.}\ \bibnamefont
  {Barrett}}, \bibinfo {author} {\bibfnamefont {H.}~\bibnamefont {Garcke}}, \
  and\ \bibinfo {author} {\bibfnamefont {R.}~\bibnamefont {N{\"u}rnberg}},\
  }\href@noop {} {\bibfield  {journal} {\bibinfo  {journal} {Physical Review
  E}\ }\textbf {\bibinfo {volume} {86}},\ \bibinfo {pages} {011604} (\bibinfo
  {year} {2012})}\BibitemShut {NoStop}%
\bibitem [{\citenamefont {Libbrecht}(2006)}]{libbrecht2006ken}%
  \BibitemOpen
  \bibfield  {author} {\bibinfo {author} {\bibfnamefont {K.~G.}\ \bibnamefont
  {Libbrecht}},\ }\href@noop {} {\emph {\bibinfo {title} {Ken Libbrecht's field
  guide to snowflakes}}}\ (\bibinfo  {publisher} {Voyageur Press},\ \bibinfo
  {year} {2006})\BibitemShut {NoStop}%
\bibitem [{\citenamefont {Furukawa}\ and\ \citenamefont
  {Shimada}(1993)}]{furukawa1993three}%
  \BibitemOpen
  \bibfield  {author} {\bibinfo {author} {\bibfnamefont {Y.}~\bibnamefont
  {Furukawa}}\ and\ \bibinfo {author} {\bibfnamefont {W.}~\bibnamefont
  {Shimada}},\ }\href@noop {} {\bibfield  {journal} {\bibinfo  {journal}
  {Journal of crystal growth}\ }\textbf {\bibinfo {volume} {128}},\ \bibinfo
  {pages} {234} (\bibinfo {year} {1993})}\BibitemShut {NoStop}%
\bibitem [{\citenamefont {Saito}(1996)}]{saito1996statistical}%
  \BibitemOpen
  \bibfield  {author} {\bibinfo {author} {\bibfnamefont {Y.}~\bibnamefont
  {Saito}},\ }\href@noop {} {\emph {\bibinfo {title} {Statistical physics of
  crystal growth}}},\ Vol.~\bibinfo {volume} {2}\ (\bibinfo  {publisher} {World
  Scientific},\ \bibinfo {year} {1996})\BibitemShut {NoStop}%
\bibitem [{\citenamefont {Sazaki}\ \emph {et~al.}(2010)\citenamefont {Sazaki},
  \citenamefont {Zepeda}, \citenamefont {Nakatsubo}, \citenamefont {Yokoyama},\
  and\ \citenamefont {Furukawa}}]{sazaki2010elementary}%
  \BibitemOpen
  \bibfield  {author} {\bibinfo {author} {\bibfnamefont {G.}~\bibnamefont
  {Sazaki}}, \bibinfo {author} {\bibfnamefont {S.}~\bibnamefont {Zepeda}},
  \bibinfo {author} {\bibfnamefont {S.}~\bibnamefont {Nakatsubo}}, \bibinfo
  {author} {\bibfnamefont {E.}~\bibnamefont {Yokoyama}}, \ and\ \bibinfo
  {author} {\bibfnamefont {Y.}~\bibnamefont {Furukawa}},\ }\href@noop {}
  {\bibfield  {journal} {\bibinfo  {journal} {Proceedings of the National
  Academy of Sciences}\ }\textbf {\bibinfo {volume} {107}},\ \bibinfo {pages}
  {19702} (\bibinfo {year} {2010})}\BibitemShut {NoStop}%
\bibitem [{\citenamefont {Kikuchi}\ and\ \citenamefont
  {Uyeda}(1998)}]{Kikuchi1998169}%
  \BibitemOpen
  \bibfield  {author} {\bibinfo {author} {\bibfnamefont {K.}~\bibnamefont
  {Kikuchi}}\ and\ \bibinfo {author} {\bibfnamefont {H.}~\bibnamefont
  {Uyeda}},\ }\href@noop {} {\bibfield  {journal} {\bibinfo  {journal}
  {Atmospheric Research}\ }\textbf {\bibinfo {volume} {47–48}},\ \bibinfo
  {pages} {169 } (\bibinfo {year} {1998})}\BibitemShut {NoStop}%
\bibitem [{\citenamefont {Singer-Loginova}\ and\ \citenamefont
  {Singer}(2008)}]{singer2008phase}%
  \BibitemOpen
  \bibfield  {author} {\bibinfo {author} {\bibfnamefont {I.}~\bibnamefont
  {Singer-Loginova}}\ and\ \bibinfo {author} {\bibfnamefont {H.}~\bibnamefont
  {Singer}},\ }\href@noop {} {\bibfield  {journal} {\bibinfo  {journal}
  {Reports on progress in physics}\ }\textbf {\bibinfo {volume} {71}},\
  \bibinfo {pages} {106501} (\bibinfo {year} {2008})}\BibitemShut {NoStop}%
\bibitem [{\citenamefont {Karma}\ and\ \citenamefont
  {Rappel}(1996)}]{karma1996phase}%
  \BibitemOpen
  \bibfield  {author} {\bibinfo {author} {\bibfnamefont {A.}~\bibnamefont
  {Karma}}\ and\ \bibinfo {author} {\bibfnamefont {W.-J.}\ \bibnamefont
  {Rappel}},\ }\href@noop {} {\bibfield  {journal} {\bibinfo  {journal}
  {Physical Review E}\ }\textbf {\bibinfo {volume} {53}},\ \bibinfo {pages}
  {R3017} (\bibinfo {year} {1996})}\BibitemShut {NoStop}%
\bibitem [{\citenamefont {Ramirez}\ \emph {et~al.}(2004)\citenamefont
  {Ramirez}, \citenamefont {Beckermann}, \citenamefont {Karma},\ and\
  \citenamefont {Diepers}}]{ramirez2004phase}%
  \BibitemOpen
  \bibfield  {author} {\bibinfo {author} {\bibfnamefont {J.}~\bibnamefont
  {Ramirez}}, \bibinfo {author} {\bibfnamefont {C.}~\bibnamefont {Beckermann}},
  \bibinfo {author} {\bibfnamefont {A.}~\bibnamefont {Karma}}, \ and\ \bibinfo
  {author} {\bibfnamefont {H.-J.}\ \bibnamefont {Diepers}},\ }\href@noop {}
  {\bibfield  {journal} {\bibinfo  {journal} {Physical Review E}\ }\textbf
  {\bibinfo {volume} {69}},\ \bibinfo {pages} {051607} (\bibinfo {year}
  {2004})}\BibitemShut {NoStop}%
\bibitem [{\citenamefont {Cartalade}\ \emph {et~al.}(2016)\citenamefont
  {Cartalade}, \citenamefont {Younsi},\ and\ \citenamefont
  {Plapp}}]{cartalade2016lattice}%
  \BibitemOpen
  \bibfield  {author} {\bibinfo {author} {\bibfnamefont {A.}~\bibnamefont
  {Cartalade}}, \bibinfo {author} {\bibfnamefont {A.}~\bibnamefont {Younsi}}, \
  and\ \bibinfo {author} {\bibfnamefont {M.}~\bibnamefont {Plapp}},\
  }\href@noop {} {\bibfield  {journal} {\bibinfo  {journal} {Computers \&
  Mathematics with Applications}\ }\textbf {\bibinfo {volume} {71}},\ \bibinfo
  {pages} {1784} (\bibinfo {year} {2016})}\BibitemShut {NoStop}%
\bibitem [{\citenamefont {Barrett}\ \emph {et~al.}(2013)\citenamefont
  {Barrett}, \citenamefont {Garcke},\ and\ \citenamefont
  {N{\"u}rnberg}}]{barrett2013stable}%
  \BibitemOpen
  \bibfield  {author} {\bibinfo {author} {\bibfnamefont {J.~W.}\ \bibnamefont
  {Barrett}}, \bibinfo {author} {\bibfnamefont {H.}~\bibnamefont {Garcke}}, \
  and\ \bibinfo {author} {\bibfnamefont {R.}~\bibnamefont {N{\"u}rnberg}},\
  }\href@noop {} {\bibfield  {journal} {\bibinfo  {journal} {ZAMM-Journal of
  Applied Mathematics and Mechanics/Zeitschrift f{\"u}r Angewandte Mathematik
  und Mechanik}\ }\textbf {\bibinfo {volume} {93}},\ \bibinfo {pages} {719}
  (\bibinfo {year} {2013})}\BibitemShut {NoStop}%
\bibitem [{\citenamefont {Kobayashi}(1993)}]{kobayashi1993modeling}%
  \BibitemOpen
  \bibfield  {author} {\bibinfo {author} {\bibfnamefont {R.}~\bibnamefont
  {Kobayashi}},\ }\href@noop {} {\bibfield  {journal} {\bibinfo  {journal}
  {Physica D: Nonlinear Phenomena}\ }\textbf {\bibinfo {volume} {63}},\
  \bibinfo {pages} {410} (\bibinfo {year} {1993})}\BibitemShut {NoStop}%
\bibitem [{\citenamefont {Debierre}\ \emph {et~al.}(2003)\citenamefont
  {Debierre}, \citenamefont {Karma}, \citenamefont {Celestini},\ and\
  \citenamefont {Gu{\'e}rin}}]{debierre2003phase}%
  \BibitemOpen
  \bibfield  {author} {\bibinfo {author} {\bibfnamefont {J.-M.}\ \bibnamefont
  {Debierre}}, \bibinfo {author} {\bibfnamefont {A.}~\bibnamefont {Karma}},
  \bibinfo {author} {\bibfnamefont {F.}~\bibnamefont {Celestini}}, \ and\
  \bibinfo {author} {\bibfnamefont {R.}~\bibnamefont {Gu{\'e}rin}},\
  }\href@noop {} {\bibfield  {journal} {\bibinfo  {journal} {Physical Review
  E}\ }\textbf {\bibinfo {volume} {68}},\ \bibinfo {pages} {041604} (\bibinfo
  {year} {2003})}\BibitemShut {NoStop}%
\bibitem [{\citenamefont {Eggleston}\ \emph {et~al.}(2001)\citenamefont
  {Eggleston}, \citenamefont {McFadden},\ and\ \citenamefont
  {Voorhees}}]{eggleston2001phase}%
  \BibitemOpen
  \bibfield  {author} {\bibinfo {author} {\bibfnamefont {J.~J.}\ \bibnamefont
  {Eggleston}}, \bibinfo {author} {\bibfnamefont {G.~B.}\ \bibnamefont
  {McFadden}}, \ and\ \bibinfo {author} {\bibfnamefont {P.~W.}\ \bibnamefont
  {Voorhees}},\ }\href@noop {} {\bibfield  {journal} {\bibinfo  {journal}
  {Physica D: Nonlinear Phenomena}\ }\textbf {\bibinfo {volume} {150}},\
  \bibinfo {pages} {91} (\bibinfo {year} {2001})}\BibitemShut {NoStop}%
\bibitem [{\citenamefont {Libbrecht}(2012)}]{libbrecht2012toward}%
  \BibitemOpen
  \bibfield  {author} {\bibinfo {author} {\bibfnamefont {K.~G.}\ \bibnamefont
  {Libbrecht}},\ }\href@noop {} {\bibfield  {journal} {\bibinfo  {journal}
  {arXiv preprint arXiv:1211.5555}\ } (\bibinfo {year} {2012})}\BibitemShut
  {NoStop}%
\bibitem [{\citenamefont {Langer}\ \emph {et~al.}(1978)\citenamefont {Langer},
  \citenamefont {Sekerka},\ and\ \citenamefont {Fujioka}}]{langer1978evidence}%
  \BibitemOpen
  \bibfield  {author} {\bibinfo {author} {\bibfnamefont {J.}~\bibnamefont
  {Langer}}, \bibinfo {author} {\bibfnamefont {R.}~\bibnamefont {Sekerka}}, \
  and\ \bibinfo {author} {\bibfnamefont {T.}~\bibnamefont {Fujioka}},\
  }\href@noop {} {\bibfield  {journal} {\bibinfo  {journal} {Journal of Crystal
  Growth}\ }\textbf {\bibinfo {volume} {44}},\ \bibinfo {pages} {414} (\bibinfo
  {year} {1978})}\BibitemShut {NoStop}%
\bibitem [{\citenamefont {Langer}\ and\ \citenamefont
  {M{\"u}ller-Krumbhaar}(1978)}]{langer1978theory}%
  \BibitemOpen
  \bibfield  {author} {\bibinfo {author} {\bibfnamefont {J.}~\bibnamefont
  {Langer}}\ and\ \bibinfo {author} {\bibfnamefont {H.}~\bibnamefont
  {M{\"u}ller-Krumbhaar}},\ }\href@noop {} {\bibfield  {journal} {\bibinfo
  {journal} {Acta Metallurgica}\ }\textbf {\bibinfo {volume} {26}},\ \bibinfo
  {pages} {1681} (\bibinfo {year} {1978})}\BibitemShut {NoStop}%
\bibitem [{\citenamefont {Amar}\ and\ \citenamefont
  {Brener}(1993)}]{amar1993theory}%
  \BibitemOpen
  \bibfield  {author} {\bibinfo {author} {\bibfnamefont {M.~B.}\ \bibnamefont
  {Amar}}\ and\ \bibinfo {author} {\bibfnamefont {E.}~\bibnamefont {Brener}},\
  }\href@noop {} {\bibfield  {journal} {\bibinfo  {journal} {Physical review
  letters}\ }\textbf {\bibinfo {volume} {71}},\ \bibinfo {pages} {589}
  (\bibinfo {year} {1993})}\BibitemShut {NoStop}%
\bibitem [{\citenamefont {Brener}(1996)}]{brener1996three}%
  \BibitemOpen
  \bibfield  {author} {\bibinfo {author} {\bibfnamefont {E.}~\bibnamefont
  {Brener}},\ }\href@noop {} {\bibfield  {journal} {\bibinfo  {journal}
  {Journal of crystal growth}\ }\textbf {\bibinfo {volume} {166}},\ \bibinfo
  {pages} {339} (\bibinfo {year} {1996})}\BibitemShut {NoStop}%
\bibitem [{\citenamefont {Libbrecht}\ \emph {et~al.}(2002)\citenamefont
  {Libbrecht}, \citenamefont {Crosby},\ and\ \citenamefont
  {Swanson}}]{libbrecht2002electrically}%
  \BibitemOpen
  \bibfield  {author} {\bibinfo {author} {\bibfnamefont {K.~G.}\ \bibnamefont
  {Libbrecht}}, \bibinfo {author} {\bibfnamefont {T.}~\bibnamefont {Crosby}}, \
  and\ \bibinfo {author} {\bibfnamefont {M.}~\bibnamefont {Swanson}},\
  }\href@noop {} {\bibfield  {journal} {\bibinfo  {journal} {Journal of crystal
  growth}\ }\textbf {\bibinfo {volume} {240}},\ \bibinfo {pages} {241}
  (\bibinfo {year} {2002})}\BibitemShut {NoStop}%
\bibitem [{\citenamefont {Libbrecht}()}]{siteLibbrecht}%
  \BibitemOpen
  \bibfield  {author} {\bibinfo {author} {\bibfnamefont {K.~G.}\ \bibnamefont
  {Libbrecht}},\ }\href@noop {} {\enquote {\bibinfo {title}
  {Snowcrystals.com},}\ }\bibinfo {howpublished}
  {\url{http://www.snowcrystals.com}}\BibitemShut {NoStop}%
\bibitem [{\citenamefont {Mullins}\ and\ \citenamefont
  {Sekerka}(1964)}]{mullins1964stability}%
  \BibitemOpen
  \bibfield  {author} {\bibinfo {author} {\bibfnamefont {W.~W.}\ \bibnamefont
  {Mullins}}\ and\ \bibinfo {author} {\bibfnamefont {R.}~\bibnamefont
  {Sekerka}},\ }\href@noop {} {\bibfield  {journal} {\bibinfo  {journal}
  {Journal of applied physics}\ }\textbf {\bibinfo {volume} {35}},\ \bibinfo
  {pages} {444} (\bibinfo {year} {1964})}\BibitemShut {NoStop}%
\bibitem [{\citenamefont {Berg}(1938)}]{berg1938crystal}%
  \BibitemOpen
  \bibfield  {author} {\bibinfo {author} {\bibfnamefont {W.}~\bibnamefont
  {Berg}},\ }in\ \href@noop {} {\emph {\bibinfo {booktitle} {Proceedings of the
  Royal Society of London A: Mathematical, Physical and Engineering
  Sciences}}},\ Vol.\ \bibinfo {volume} {164}\ (\bibinfo {organization} {The
  Royal Society},\ \bibinfo {year} {1938})\ pp.\ \bibinfo {pages}
  {79--95}\BibitemShut {NoStop}%
\bibitem [{\citenamefont {Fujioka}\ and\ \citenamefont
  {Sekerka}(1974)}]{fujioka1974morphological}%
  \BibitemOpen
  \bibfield  {author} {\bibinfo {author} {\bibfnamefont {T.}~\bibnamefont
  {Fujioka}}\ and\ \bibinfo {author} {\bibfnamefont {R.}~\bibnamefont
  {Sekerka}},\ }\href@noop {} {\bibfield  {journal} {\bibinfo  {journal}
  {Journal of Crystal Growth}\ }\textbf {\bibinfo {volume} {24}},\ \bibinfo
  {pages} {84} (\bibinfo {year} {1974})}\BibitemShut {NoStop}%
\bibitem [{\citenamefont {Singer}\ and\ \citenamefont
  {Bilgram}(2004)}]{singer2004three}%
  \BibitemOpen
  \bibfield  {author} {\bibinfo {author} {\bibfnamefont {H.}~\bibnamefont
  {Singer}}\ and\ \bibinfo {author} {\bibfnamefont {J.}~\bibnamefont
  {Bilgram}},\ }\href@noop {} {\bibfield  {journal} {\bibinfo  {journal} {EPL
  (Europhysics Letters)}\ }\textbf {\bibinfo {volume} {68}},\ \bibinfo {pages}
  {240} (\bibinfo {year} {2004})}\BibitemShut {NoStop}%
\bibitem [{\citenamefont {Libbrecht}(2016)}]{libbrecht2016toward}%
  \BibitemOpen
  \bibfield  {author} {\bibinfo {author} {\bibfnamefont {K.~G.}\ \bibnamefont
  {Libbrecht}},\ }\href@noop {} {\bibfield  {journal} {\bibinfo  {journal}
  {arXiv preprint arXiv:1602.08528}\ } (\bibinfo {year} {2016})}\BibitemShut
  {NoStop}%
\bibitem [{\citenamefont {Almgren}\ \emph {et~al.}(1993)\citenamefont
  {Almgren}, \citenamefont {Dai},\ and\ \citenamefont
  {Hakim}}]{almgren1993scaling}%
  \BibitemOpen
  \bibfield  {author} {\bibinfo {author} {\bibfnamefont {R.}~\bibnamefont
  {Almgren}}, \bibinfo {author} {\bibfnamefont {W.-S.}\ \bibnamefont {Dai}}, \
  and\ \bibinfo {author} {\bibfnamefont {V.}~\bibnamefont {Hakim}},\
  }\href@noop {} {\bibfield  {journal} {\bibinfo  {journal} {Physical review
  letters}\ }\textbf {\bibinfo {volume} {71}},\ \bibinfo {pages} {3461}
  (\bibinfo {year} {1993})}\BibitemShut {NoStop}%
\bibitem [{\citenamefont {Zener}(1949)}]{zener1949theory}%
  \BibitemOpen
  \bibfield  {author} {\bibinfo {author} {\bibfnamefont {C.}~\bibnamefont
  {Zener}},\ }\href@noop {} {\bibfield  {journal} {\bibinfo  {journal} {Journal
  of Applied Physics}\ }\textbf {\bibinfo {volume} {20}},\ \bibinfo {pages}
  {950} (\bibinfo {year} {1949})}\BibitemShut {NoStop}%
\bibitem [{\citenamefont {Libbrecht}(2008)}]{libbrecht2008crystal}%
  \BibitemOpen
  \bibfield  {author} {\bibinfo {author} {\bibfnamefont {K.~G.}\ \bibnamefont
  {Libbrecht}},\ }\href@noop {} {\bibfield  {journal} {\bibinfo  {journal}
  {arXiv preprint arXiv:0810.0689}\ } (\bibinfo {year} {2008})}\BibitemShut
  {NoStop}%
\bibitem [{\citenamefont {Marshall}\ and\ \citenamefont
  {Langleben}(1954)}]{marshall1954theory}%
  \BibitemOpen
  \bibfield  {author} {\bibinfo {author} {\bibfnamefont {J.~S.}\ \bibnamefont
  {Marshall}}\ and\ \bibinfo {author} {\bibfnamefont {M.~P.}\ \bibnamefont
  {Langleben}},\ }\href@noop {} {\bibfield  {journal} {\bibinfo  {journal}
  {Journal of Meteorology}\ }\textbf {\bibinfo {volume} {11}},\ \bibinfo
  {pages} {104} (\bibinfo {year} {1954})}\BibitemShut {NoStop}%
\bibitem [{\citenamefont {Jeong}\ \emph {et~al.}(2001)\citenamefont {Jeong},
  \citenamefont {Goldenfeld},\ and\ \citenamefont
  {Dantzig}}]{PhysRevE.64.041602}%
  \BibitemOpen
  \bibfield  {author} {\bibinfo {author} {\bibfnamefont {J.-H.}\ \bibnamefont
  {Jeong}}, \bibinfo {author} {\bibfnamefont {N.}~\bibnamefont {Goldenfeld}}, \
  and\ \bibinfo {author} {\bibfnamefont {J.~A.}\ \bibnamefont {Dantzig}},\
  }\href {\doibase 10.1103/PhysRevE.64.041602} {\bibfield  {journal} {\bibinfo
  {journal} {Phys. Rev. E}\ }\textbf {\bibinfo {volume} {64}},\ \bibinfo
  {pages} {041602} (\bibinfo {year} {2001})}\BibitemShut {NoStop}%
\bibitem [{\citenamefont {Whiteway}\ \emph {et~al.}(2015)\citenamefont
  {Whiteway}, \citenamefont {Yang}, \citenamefont {Yu},\ and\ \citenamefont
  {Hilke}}]{whiteway2015time}%
  \BibitemOpen
  \bibfield  {author} {\bibinfo {author} {\bibfnamefont {E.}~\bibnamefont
  {Whiteway}}, \bibinfo {author} {\bibfnamefont {W.}~\bibnamefont {Yang}},
  \bibinfo {author} {\bibfnamefont {V.}~\bibnamefont {Yu}}, \ and\ \bibinfo
  {author} {\bibfnamefont {M.}~\bibnamefont {Hilke}},\ }\href@noop {}
  {\bibfield  {journal} {\bibinfo  {journal} {arXiv preprint arXiv:1509.01579}\
  } (\bibinfo {year} {2015})}\BibitemShut {NoStop}%
\end{thebibliography}
%


\appendix

\section{comparison of simulations with real snowflakes}
\label{annexe1}

\begin{figure}[H]
\begin{subfigmatrix}{2}
\subfigure[~~Simple star  from \cite{siteLibbrecht}]{\includegraphics[height=3.8cm]{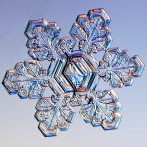}\label{fig:photo_star}}
\subfigure[~~Simple star]{\includegraphics[height=3.8cm]{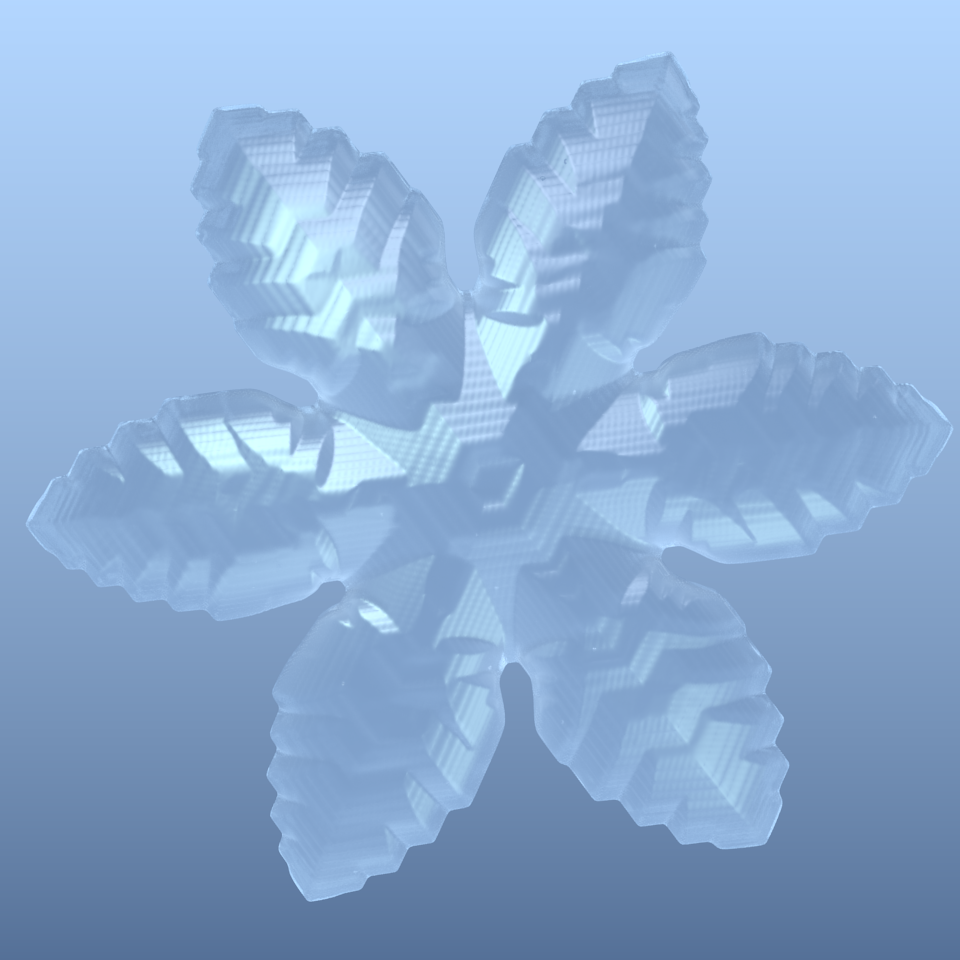}\label{fig:simu_star}}
\subfigure[~~Stellar dendrite  from \cite{siteLibbrecht}]{\includegraphics[height=3.8cm]{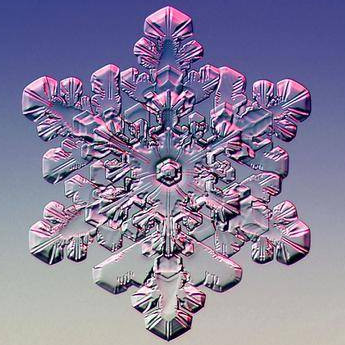}\label{fig:photo_stellar}}
\subfigure[~~Stellar dendrite]{\includegraphics[height=3.8cm]{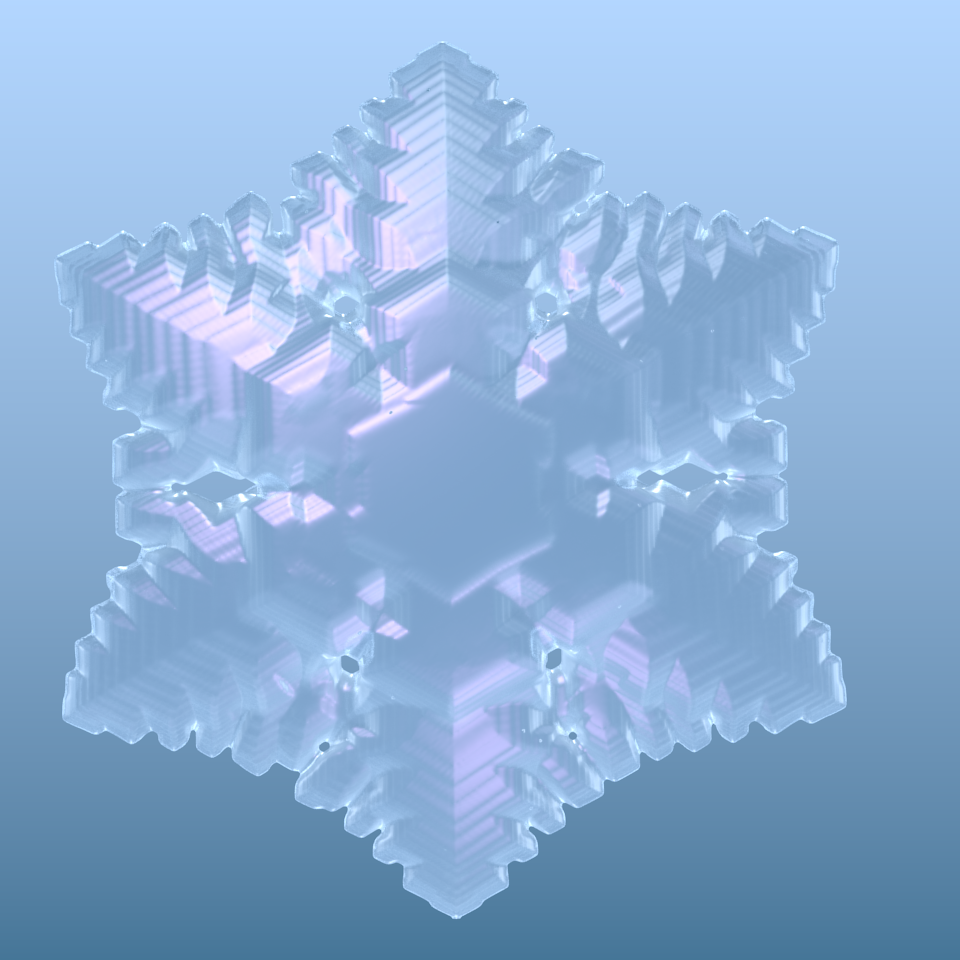}\label{fig:simu_stellar}}\vspace{-0.3cm}
\subfigure[~~Sectored plate from \cite{siteLibbrecht}]{\includegraphics[height=3.8cm]{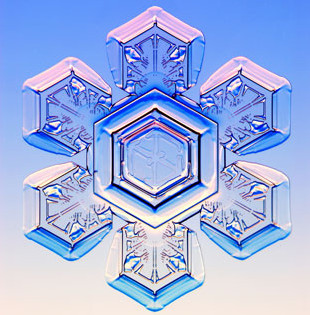}\label{fig:photo_sector}}
\subfigure[~~Sectored plate]{\includegraphics[height=3.8cm]{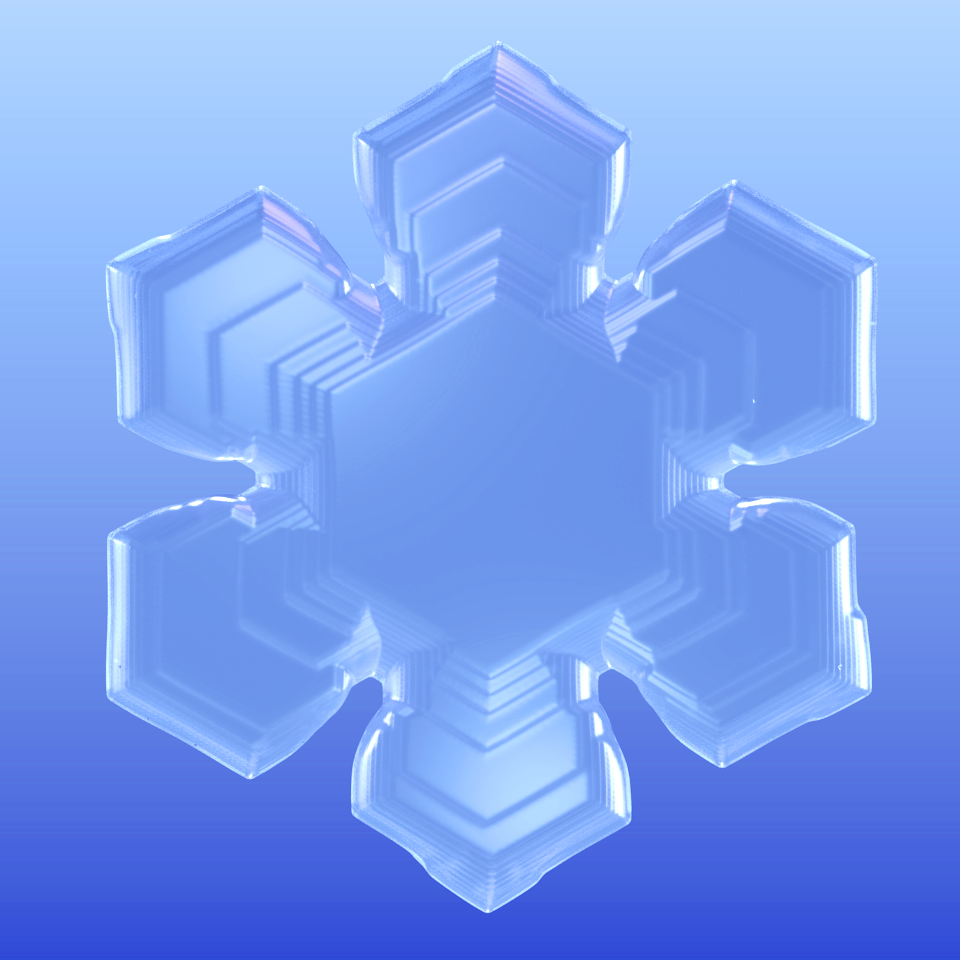}\label{fig:simu_sector}}\vspace{-0.3cm}
\end{subfigmatrix}
\caption{Comparison between real snowflakes photographs (left) taken from \cite{siteLibbrecht}, and our phase field simulations (right). Visual rendering for our simulations uses the software Blender.}
\label{fig:comparaison} 
\end{figure} 

\section{simulation method}
\label{annexe2}

Phase field simulations were performed using the Fourier-spectral semi-implicit scheme with periodic boundary conditions. For horizontal growth, simulations were performed with grid spacing $\Delta x=\Delta y=\Delta z=0.8$, and time step $\Delta t=0.05$ \cite{ramirez2004phase}, on a 400 $\times$ 400 $\times$ 64 simulation box. Vertical growth simulations required a more precise discretization. Grid spacing was thus reduced to $\Delta x=\Delta y=\Delta z=0.4$, and $\Delta t=0.01$ as in \cite{ramirez2004phase}, on a 128 $\times$ 128 $\times$ 256 simulation box. Snow crystal growth was initiated by a circular-disk shape germ ($\phi=1$) of radius $R=8\Delta x$, within water vapour ($\phi=-1$) of homogeneous reduced supersaturation $u_0>0$. A new regularization method for high interfacial energy leading to crystal missing orientations was also adapted from two dimensions \cite{eggleston2001phase}, to three dimensions. This allowed to overcome the restriction for the critical values of the anisotropy constants and choose $\epsilon_{xy}>1/35$ and $\epsilon_{z}>1/3$, as required to achieve faceting \cite{singer2008phase}.

\section{3D faceting algorithm}
\label{annexe3}

Faceting requires the anisotropy constants $\epsilon_{xy}$ and $\epsilon_{z}$ values in $A(\vect{n})$, to exceed $1/35$ for 6-fold horizontal symmetry, and $1/3$ for 2-fold vertical symmetry \cite{singer2008phase}. Above such values, metastable and unstable crystal orientations $\vect{n}$ where the stiffness becomes negative, emerge. $A(\vect{n})$ must hence be modified, so that the stiffness remains positive. A preliminary step is to determine missing orientations. This can be done by detecting sign change in the stiffness. This causes a local convexity inversion in the 2D polar plot of  $1/A(\theta,\phi_0)$, for any fixed azimuthal angle $\phi_0$,  or in the polar plot of  $1/A(\theta_0,\phi)$ for any fixed polar angle $\theta_0$. This reads $A+\partial^2_{\theta\theta} A=0$ and $A+\partial^2_{\phi\phi} A=0$ respectively, providing us with the maximum anisotropy constants $\epsilon_{xy}^m$ and $\epsilon_{z}^m$ before missing orientation:

\begin{equation}
\left\{
\begin{aligned}
&\epsilon_{xy}^m=\frac{1+\epsilon_z \cos(2\phi)}{35}\\
&\epsilon_{z}^m=\frac{1+\epsilon_{xy} \cos(6\theta)}{3}.
\end{aligned}
\right.
\label{eps}
\end{equation}
Restricting the study to $\theta \in [-\pi/6,\pi/6]$ and $\phi \in [0,\pi]$ due to rotation periodicity of $A(\theta,\phi)$, missing $\theta$ orientations lie within a $\phi$-dependant angular sector $[-\theta^m(\phi)+k\pi/3,\theta^m(\phi)+k\pi/3]$ ($k\in \mathbb{Z}$) for $\epsilon_{xy}>\epsilon_{xy}^m$. Same goes for $\phi$ orientations lying within $[-\phi^m(\theta)+k\pi,\phi^m(\theta)+k\pi]$ for $\epsilon_z>\epsilon_{z}^m$. Following the procedure developed by Eggleston in \cite{eggleston2001phase}, $\theta^m$ and $\phi^m$ respectively satisfy $\partial_{\theta}[\cos(\theta)/A(\theta,\phi)]=0$ and $\partial_{\phi}[\cos(\phi)/A(\theta,\phi)]=0$, leading to:

\begin{equation}
\left\{
\begin{aligned}
&\frac{6\epsilon_{xy} \sin(6\theta^m) \cos(\theta^m)}{\sin(\theta^m)}=\left(1+\epsilon_{xy}\cos(6\theta^m)+\epsilon_z cos(2\phi)\right)\\
&\frac{2\epsilon_{z} \sin(2\phi^m) \cos(\phi^m)}{\sin(\phi^m)}=\left(1+\epsilon_{xy}\cos(6\theta)+\epsilon_z cos(2\phi^m)\right).
\end{aligned}
\right.
\label{angles}
\end{equation}
The \emph{regularization} step finally consists in replacing $A(\theta,\phi)$ by a simple trigonometric function for missing orientation, as suggested by Debierre et al. \cite{debierre2003phase}:

\begin{equation}
A(\vect{n})=
\left\{
\begin{aligned}
&A^{\theta}(\theta,\phi)=A_1(\phi)+B_1\cos(\theta), \text{ } |\theta|<\theta^m,  |\phi|\geq \phi^m\\
& A^{\phi}(\theta,\phi)=A_2(\theta)+B_2\cos(\phi), \text{ } |\theta|\geq\theta^m,  |\phi|< \phi^m\\
&\alpha(\theta,\phi) A^{\theta} +(1-\alpha(\theta,\phi)) A^{\phi}, \text{ } |\theta|<\theta^m,  |\phi|<\phi^m \\
&1+\epsilon_{xy}\cos(6\theta)+\epsilon_{z}\cos(2\psi) \text{ } \text{otherwise}.
\end{aligned}
\right.
\label{Aregul}
\end{equation}
Coefficients $A_1(\phi)$, $B_1$, $A_2(\theta)$, $B_2$ and $\alpha(\theta,\phi)$ are set to ensure continuity of $A(\vect{n})$ and its derivatives:

\begin{equation}
\begin{aligned}
&A_1(\phi)=1+\epsilon_{xy} \cos(6\theta^m)+\epsilon_z\cos(2\phi)-\frac{6\epsilon_{xy} \sin(6\theta^m)}{\sin(\theta^m)}\cos(\theta^m)\\
&B_1=\frac{6\epsilon_{xy} \sin(6\theta^m)}{\sin(\theta^m)}\\
&A_2(\theta)=1+\epsilon_{xy} \cos(6\theta)+\epsilon_z\cos(2\phi^m)-\frac{2\epsilon_{z} \sin(2\phi^m)}{\sin(\phi^m)}\cos(\phi^m)\\
&B_2=\frac{2\epsilon_{z} \sin(2\phi^m)}{\sin(\phi^m)}\\
&\alpha(\theta,\phi)=\frac{|\theta-\theta^m|}{\sqrt{(\theta-\theta^m)^2+(\phi-\phi^m)^2}}.
\end{aligned}
\label{coefs}
\end{equation}

\end{document}